\documentclass[useAMS,usenatbib]{mn2e}
\usepackage{graphicx}
\usepackage{epstopdf}
\usepackage{amsmath}
\title[]{How does an asymmetric magnetic field change the vertical structure of a hot accretion flow? }
\author[ Samadi  M.; Abbassi S.; Lovelace R. V. E ]{
  Samadi M.$^{1}$\thanks{$m _-samadi_-m$@yahoo.com}; Abbassi S.$^{1,2}$\thanks{abbassi@um.ac.ir}; 
    Lovelace R. V. E.$^{3}$\thanks{E-mail:lovelace@astro.cornell.edu}  
\\
$^{1}$Department of Physics, School of Sciences, Ferdowsi University of Mashhad, Mashhad, 91775-1436, Iran\\
$^{2}$School of Astronomy, Institute for Studies in
Theoretical Physics and Mathematics, P.O.Box 19395-5531, Tehran,Iran\\
$^{3}$Department of Astronomy, Cornell University, Ithaca, NY 14853 
}

\date{}
\begin{document}
\pagerange{\pageref{firstpage}--\pageref{lastpage}} \pubyear{2012}

\maketitle \label{firstpage}

\begin{abstract}
This paper explores the effects of large-scale magnetic fields in hot accretion flows for asymmetric configurations  with respect to the equatorial plane. The solutions that we have found show that the large-scale asymmetric magnetic field can significantly affect the dynamics of the flow and also cause notable outflows in the outer parts. Previously, we treated a viscous resistive accreting disc in the presence of an odd symmetric $\textbf{B-}$ field about the equatorial plane. Now we extend our earlier work by taking into account another configuration of large-scale magnetic field which is no longer symmetric. We provide asymmetric field structures with small deviations from even and odd symmetric $\textbf{B-}$field. Our results show that the disc's dynamics and appearance become different above and below the equatorial plane. The set of solutions also predicts that even a small deviation in a symmetric field causes the disc to compress on one side and expand on the other.  In some cases, our solution represents a very strong outflow from just one side of the disc. Therefore, the solution may potentially explain the origin of one-sided jets in radio galaxies.
\end{abstract}

\section{INTRODUCTION}
To investigate outstanding phenomena happening in celestial objects such as black holes, neutron stars and close binary stars it is common to apply 
accretion disc models.  The main idea is heating gas due to MHD turbulence acting as an effective viscosity (Abramowicz et al. 2013). Then this extra energy may stay in or leave the disc as a radiation as in the standard model of Shakura \& Sunyaev (1973). 
   In the optically thin plasma and in the case of sub-Eddington luminosity, energy is advected and enhances the whole entropy of the system (which called advection-dominated accretion flow or ADAF). Thus the flow becomes hot and thick. This  approach was applied for the first time by Ichimaru (1977) to explain the transition between low and high states in X-ray in the spectrum of Cyg X-1. Narayan \& Yi (1994, 1995 a, b) established a new framework to study this branch of accreting systems. They assumed that the energy which is trapped in the flow is a certain fraction of viscous dissipation. During the past decade intensive studies have been performed on black hole hot accretion flows (see Yuan \& Narayan 2014, for a review). It is now known that hot accretion flow is quite common in the universe and, as a model, is able to reproduce power spectrum of low-luminosity active galactic nuclei (LLAGNs), which is the majority of nearby galaxies, and the quiescent/hard states of black hole X-ray binaries as well. In order to advance our understanding of hot accretion flows, a lot of improvements have been proposed, including its multi-dimensional dynamics, disc-jet connection, radiation mechanisms, and various astrophysical applications (see Yuan \& Narayan 2014, for a review).

 One of the most important progresses associated with hot accretion flows is the discovery of strong winds launched from the disc in numerical simulations(Yuan et al. 2012 a,b; Narayan et al. 2012; Li et al. 2013; Sadowski et al. 2016, Bu et al. 2016 a, b).  
 The existence of outflow has been confirmed by the Chandra observations of the accretion flow around the super-massive black hole in the galactic centre (Wang et al. 2013). On the other hand, studying outflow/wind mechanisms is important since it plays crucial role in AGN feedback, because wind/outflow can effectively suppress the star formation (see, Ostriker et al. 2010). The presence of wind/outflow helps to have a reliable explanation for many observational features of hot accretion flows, including the spectra of black hole sources (e.g. Yuan, Quataert \& Narayan 2003), emission lines from accretion flow (e.g. Wang et al. 2013) and Fermi bubbles in the galactic centre (Mou et al. 2014). 

On the other hand, the structure of hot accretion flow is also remarkably affected by outflows, which carry huge amounts of energy, mass and momentum (Bu, Yuan \& Xie 2009; Kawabata \& Mineshige 2009; Samadi, Abbassi \& Khajavi 2014; Yuan, Wu \& Bu 2012 a,b; Bu et al. 2013; Gu 2015; Yuan et al. 2015). The study of wind/outflow was initiated by some theoretical study (Xu \& Chen 1997; Blandford \& Begelman 1999, 2004; Xue \& Wang 2005; Tanaka \& Menou 2006) and confirmed by some hydrodynamical and MHD simulations (Stone, Pringle \& Begelman 1999; Igumenshchev \& Abramowicz 1999, 2000; Stone \& Pringle 2000; Yuan, Wu \& Bu 2012 a,b). In NY95a solutions, although the meridional velocity had been ignored, the footstep of outflow was seen in positive value of Bernoulli function near the poles.  Xu \& Chen (1997) improved NY95a without neglecting $v_\theta$ and verified that gas pressure and viscosity are strong enough to initiate ejection of matters from the disc. The adiabatic inflow/outflow solution (ADIOS) model is an additional development that was suggested by Blandford \& Begelman (1999, 2004). They confirmed that the positive amount of energy in accreting gas supports outflow and formation of a wind.  Blandford \& Begelman (2004) studied another possibility to diminish accretion rate by including convectional movements of gas that appear due to captured thermal energy. However, near the surface this mechanism became inefficient and conditions were capable for outflows. Xue \& Wang (2005) extended ADIOS and didn't use hydrostatic equilibrium in the vertical direction. They could also improve NY95a with applying a limited thickness for the flow. But their solutions demonstrated just the inflow part of the disc and the presence of outflow was emphasized by adopting $r-$dependency of $\dot{M}$.  Jiao \& Wu (2011) carried on this approach to achieve solutions above and below the surface in the outflow region. They showed that outflow appears in slim discs, thin quasi-Keplerain discs as well as thick hot accretion flows. 
Furthermore, it is widely accepted that the magnetic field is an important ingredient in the dynamics of accretion flows and their emission. In particular, it is likely to be responsible for the accretion disc viscosity via magnetohydrodynamic (MHD) turbulence driven by the magnetorotational instability, as suggested by Balbus \& Hawley (1991). Although the magnetic fields play a crucial role in the dynamics of hot accretion flow, the origin, structure and strength of such magnetic fields in these flows remains unknown. To study magnetic field effects on accretion discs and to study how jets launch from them, Lovelace et al. (1987) and Wang et al. (1990) applied two different reflection symmetries about the equatorial plane. The case with odd symmetry is called dynamo-driven field and the even case is supposed to be a dipole primordial magnetic field.  Their results show that in an even field symmetry, there could be a large toroidal magnetic field inside a thin disc which may compress it in the vertical direction. Samadi et al. (2014) have studied a purely toroidal magnetic field with an odd symmetry and obtained the same effect which indicates that the flow shrinks vertically because of the magnetic field pressure. Even in the presence of the outflow which was supposed to be driven by a torodial magnetic field, the disc was seen to become thinner (Mosallanezhad et al. 2014). Dynamical effects of dipolar magnetic field with even symmetry configuration have been investigated too (Shadmehri 2004, Ghanbari et al. 2007 \& 2009). If we look for a more real case, we should consider a large-scale magnetic field with all three components. This kind of $\textbf{B}$-field has been recently studied with even and odd symmetries about the midplane in two seperate works by Mosallanezhad et al. (2016) and Samadi \& Abbassi (2016) (hereafter SA16), respectively.  

There are some considerable evidences of the existence of asymmetric jets in extragalactic radio sources such as GRO J1655-40 (Hjellming \& Rupen 1995), SS 443 (Fejes 1986, Paragi et al. 1998), Cygnus X-3 (Mioduszewski et al. 2001), M87 (Perlman et al. 1999) and M81 (Bietenholz et al. 2000). The first guess for the reason of one-sidness is the Doppler boosting effect which explains the approaching jet will appear brighter and faster than the receding one (Chagelishvili et al. 1995). Moreover, symmetrical outflows may appear asymmetric due to their orientations on the sky, particularly when they are observed through a large spatial scales with a dissipative region (Fiege \& Henriksen 1996a, b and Lery et al. 1999). Besides, some scientists believe there is an intrinsic reason. Moreover, some simulations have examined asymmetric matter ejections from stellar magnetospheres (Lovelace et al. 2010; 2014; Dyda et al. 2015). 
Barkov \& Komissarov (2010) have done a numerical simulation of asymmetric relativistic jets in close-binary gamma-ray burst systems. Fendt \& Sheikhnezami (2013) performed an axisymmetric MHD simulation of a disk-jet structure and noticed asymmetric outflows with a $10-30$ percent mass flux difference. A few studies have been theoretically focused on intrinsically asymmetric outflows such as Wang et al. (1992). They considered asymmetric magnetic field that led them to solutions with asymmetric jets. Chagelish et. al (1996) took the same configuration of the magnetic field and used results for explanation of the possible origin of one-sided jets in radio galaxies. 

In the present work, we extend our analysis of magnetic field effects to the more general case of $\textbf{B}-$field which is asymmetric about equatorial plane. We can expect that the flow will not have the reflection symmetry about the mid-plane. Therefore, outflows from top and bottom sides of disc will not be the same. We examine these possibilities in the following. The paper is organised as follows. In section 2 we present basic equations of the magnetised compressible, viscous flow. We show the self-similar solutions in section 3. The boundary conditions are introduced in section 4. The result of calculation are presented in section 5. We summarize results in section 6.

\section{Basic Equations}
To describe a plasma, we use a system of basic MHD equations in suggestions of steady state ($\partial/\partial t=0$) and axisymmetry ($\partial/\partial \phi=0$). We suppose that the flow is in advection-dominated state and has zero radiative cooling rate. We exclude  self-gravity and the general relativistic effects. The gravity force is assumed Newtonian. With these simple assumptions we consider continuity and momentum equations,
\begin{equation}
\frac{\partial\rho}{\partial t}+  \nabla \cdot (\rho\textbf{v})=0,
\end{equation}
\begin{equation}
\rho \frac{D\textbf{v}}{Dt}=-\nabla p - \rho\nabla\Phi+\textbf{F}^{\nu}+\frac{1}{4\pi}(\nabla\times\textbf{B})\times \textbf{B},
\end{equation}
where $\rho$, $p$, $\textbf{v}$ and  $\textbf{B}$  are  the density of the gas, the pressure, the time-averaged flow's velocity and the time-averaged magnetic field, respectively. The Lagrangian derivative was used, i.e. $D/Dt=\partial/\partial t+\textbf{v}\cdot\nabla$. Moreover, $\textbf{F}^\nu=-\nabla\cdot\textbf{T}^\nu$
is the viscous force, $\textbf{T}^\nu$ is the viscous stress tensor. Because of the sphere-like accretion flow, we write our equations in spherical coordinates ($r, \theta, \phi$). Therefore, large-scale magnetic field and velocity vector can be written as $\textbf{B}=B_r\hat{r}+B_\theta\hat{\theta}+B_\phi\hat{\phi}$ and $\textbf{v}=v_r\hat{r}+v_\theta\hat{\theta}+v_\phi\hat{\phi}$,  respectively. In the following, we assume that only the term $ T_{r\phi}=\rho\nu r\partial(v_\phi/r)/\partial r$   is important in the viscous stress tensor, where  $\nu$ is the kinematic viscosity. It is very common to use the $\alpha$-prescription of viscosity (Shakura and Sunyaev 1973), where the viscosity is taken in the form;   
$\nu(r,\theta)=\alpha r/(\rho v_K)[p+B^2/(8\pi)]$, 
where $\alpha$ is a dimensionless coefficient which is assumed to be a constant and independent of $r$, $v_K^2=GM/r$ is the Keplerian velocity squared and $B^2/8\pi$ is the magnetic pressure.  In the presence of magnetic field we should include the induction equation too, which is,
\begin{equation}
\frac{\partial\textbf{B}}{\partial t}=\nabla\times(\textbf{v}\times\textbf{B}-\eta\nabla\times\textbf{B}).
\end{equation}
where, $\eta$ is the magnetic diffusivity.
\begin{figure*}
\centering
\includegraphics[width=190mm]{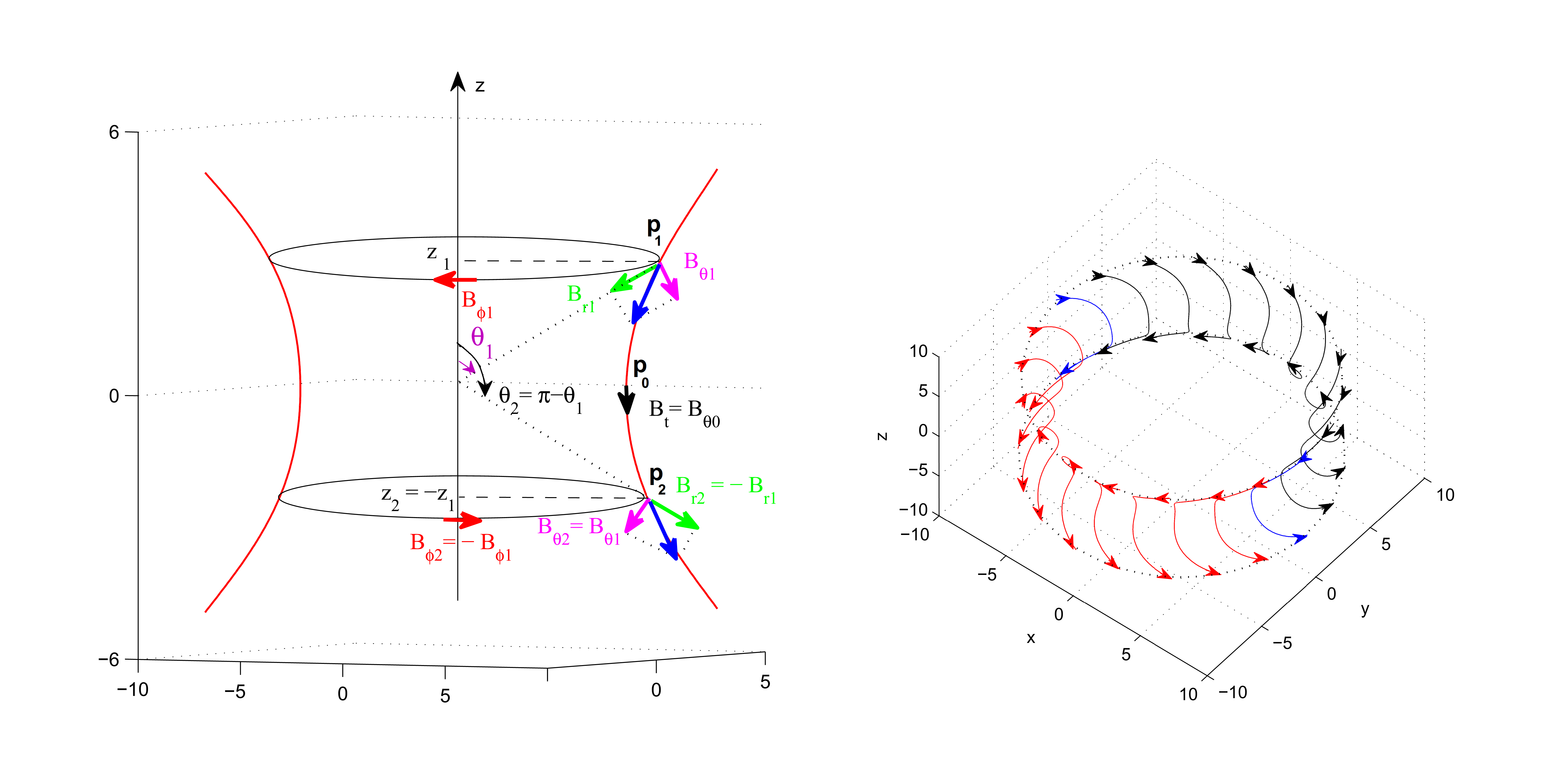}
 \caption{ Magnetic field lines with even symmetry configuration. In the left-hand side, we have plotted 2 magnetic field lines (red colour). In this panel, the poloidal components $B_r$ (green arrows) and $B_\theta$ (violet arrows) are displayed at two points $p_1$ and $p_2$. The first point is located top side at $z_1$ with the polar angle of $\theta_1<90$ above the equator and, the second point $p_2$ is below the equator with the same height and the polar angle of $\theta_2=\pi-\theta_1$. As seen, $B_r(\theta_1)=- B_r(\pi-\theta_1)$, $B_\theta(\theta_1)=B_\theta(\pi-\theta_1)$. The third point is $p_0$ with zero height and the magnetic field has only one component of $B_{\theta0}$ (black arrow). The right-hand side panel shows the same magnetic field lines in 3D space. In this panel, the reflection symmetry of $B_\phi$ is elucidated, the top arrows are clockwise and the bottom arrows are counter clockwise, i.e. $B_\phi(\theta_1)=- B_\phi(\pi-\theta_1)$. }\end{figure*}  

\begin{figure*}
\centering
\includegraphics[width=120mm]{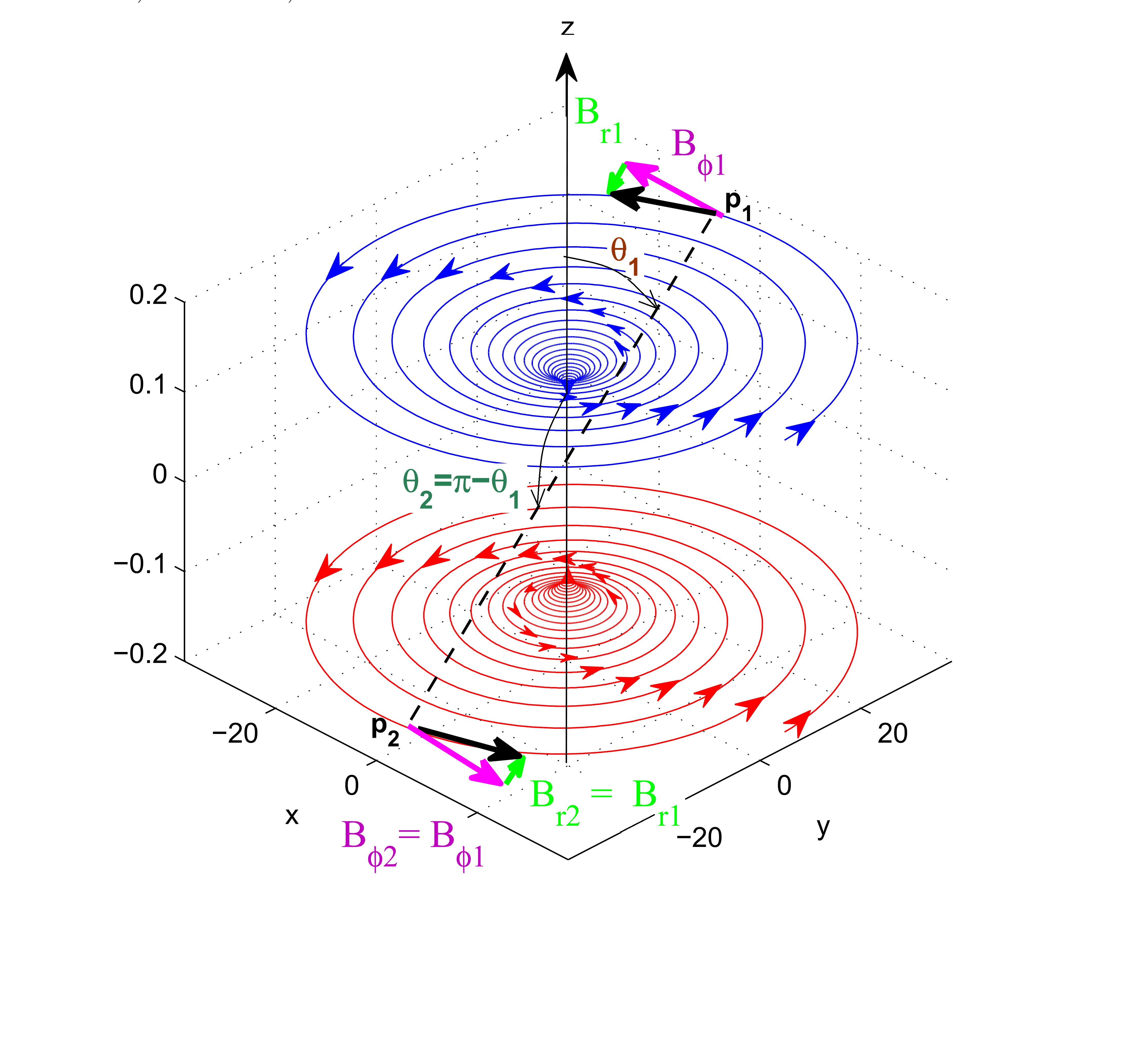}
 \caption{ Magnetic field lines with odd symmetry configuration. The magnetic field line of top side is blue and the bottom side is red. As seen the dominant component is toroidal. The two points of $p_1$ and $p_2$ are in the opposite direction of $z-$axis. $p_1$ is in $\theta_1$ polar angle and $p_2$ is in $\theta_2=\pi-\theta_1$. In these two points, the radial and toroidal components are the same, i.e. $B_r(\theta_1)=B_r(\pi-\theta_1)$, $B_\phi(\theta_1)=B_\phi(\pi-\theta_1)$. The third component of magnetic field is much smaller than $B_\phi$. Moreover, $B_{\theta0}$ (i.e. the meridional component at the equator) is exactly zero  and its sign changes throughout the equator. It might be noticed the peak of red magnetic line is upward whereas the blue one is downward.}\end{figure*}  

We should also add the equation for zero divergency of the field; $\nabla\cdot\textbf{B}=0$. Now we can rewrite $B_r$ and $B_\theta$ using the magnetic flux function $\psi$;
\begin{displaymath}
B_r=\frac{1}{r^2\sin\theta}\frac{\partial\Psi}{\partial\theta}
\end{displaymath}
\begin{displaymath}
B_\theta=-\frac{1}{r\sin\theta}\frac{\partial\Psi}{\partial r}
\end{displaymath}
 We can complete our set of equations with energy equation. Like NY94, we write the energy equation as following,
 \begin{equation}
q_+=q_{adv}+q_-,
\end{equation}
where, $q_+=T_{r\phi}r\partial/\partial r(v_\phi/r)$, $q_{adv}$ and $q_-$ are viscous heating, advective and radiative cooling, respectively. The advection energy is determined by $q_{adv}=\rho De/Dt-(p/\rho) D\rho/Dt=fq_+$,  where $e=1/(\gamma-1)p/\rho$ is entropy of accreting gas and $f$ defines the fraction of generated energy which goes towards the centre of the disc.  We write equations in the spherical coordinates and obtain 8 equations with 8 unknowns, which are density, gas pressure, 3 components of magnetic field vector and 3 ones of velocity vector. We have listed them in appendix A.

Finally, we have a set of nonlinear partial differential equations which are impossible to solve analytically. So we use a method, called self-similarity, which moderates the radial dependency of equations. 

\section{Self-Similar Solutions}
Self-similar solutions are commonly used for solving ODEs problems in fluid mechanics. As we said in the previous section, we need to use self-similar approach to remove radial dependency of quantities. So we will be able to study the vertical structure of the flow. It should be noticed that the self-similarity is unique, so we basically consider that $G$ and the central mass $M$ are held constant separately, under length, time and mass rescaling of dimensions\footnote{ This was found in the past by trial and error based on dimensional analysis, but a book 'Scale Invariance' by Henriksen gives a modern  mathematical approach. (Lery et al. 1999, Henriksen 1994). Following Lery et al. (1999) and Aburihan \& Henriksen (1999), we define all our physical quantities in terms of a fiducial radius, $r_0$ as below, }
\begin{displaymath}
\textbf{v}(r, \theta)=\sqrt{\frac{GM}{r}}\textbf{v}(\theta)
\end{displaymath}
\begin{displaymath}
\rho(r,\theta)=\frac{M}{r_0^3}\bigg( \frac{r}{r_0}\bigg)^{-n} \rho(\theta), 
\end{displaymath}
\begin{displaymath}
p(r,\theta)= \frac{GM^2}{r_0^4}\bigg(\frac{r}{r_0}\bigg)^{-(n+1)} p(\theta),
\end{displaymath}
\begin{displaymath}
\textbf{B}(r,\theta) =\sqrt{\frac{GM^2}{r_0^4}} \bigg(\frac{r}{r_0}\bigg)^{-(n+1)/2}\textbf{B}(\theta),
\end{displaymath}
\begin{equation}
\Psi(r,\theta)=\sqrt{\frac{GM^2}{r_0^4}}\bigg(\frac{r}{r_0}\bigg)^{(3-n)/2}\Psi(\theta)
\end{equation}
Substituting them to the basic equations, we obtain, 
 \begin{equation}
v_\theta\frac{d\rho}{d\theta}=-\rho\bigg[\frac{dv_\theta}{d\theta}+(\frac{3}{2}-n) v_r+\cot\theta v_\theta\bigg],
\end{equation}
\begin{displaymath}
v_\theta\frac{dv_r}{d\theta}=\frac{1}{2}v^2_r+v_\theta^2+v^2_\phi- v^2_k+(n+1)\frac{p}{\rho}\end{displaymath}
  \begin{equation}
  \hspace*{1.5cm}  +\frac{1}{8\pi\rho}\bigg[2 B_\theta\frac{dB_r}{d\theta}+(n-1)(B_\theta^2+ B_\phi^2)\bigg],
\end{equation}
 \begin{displaymath}
  v_\theta\frac{d v_\theta}{d\theta}=-\frac{1}{2}v_r v_\theta+v^2_\phi \cot\theta -\frac{1}{\rho}\frac{dp}{d\theta}
  \end{displaymath}
  \begin{equation}
\hspace{1cm}-\frac{1}{8\pi\rho}\bigg[(n-1) B_r B_{\theta}+\frac{d }{d\theta}(B_r^2+B_\phi^2)+2B_\phi^2\cot\theta\bigg],
 \end{equation}
\begin{displaymath}
v_\theta\frac{d v_\phi}{d\theta}=-\frac{1}{2}v_\phi (v_r +2 v_\theta \cot\theta)
 -(n-2)\frac{T_{r\phi}}{\rho}
 \end{displaymath}
  \begin{equation}
   \hspace{1cm}-\frac{1}{8\pi\rho }\bigg[(n-1)B_r B_\phi-2B_\theta (\frac{dB_\phi}{d\theta}+B_\phi \cot\theta)\bigg],
\end{equation}
where the above equations are the continuity and momentum equations respectively.
The energy equation becomes
\begin{equation}
(n-\frac{1}{\gamma-1})pv_r+\frac{v_\theta}{\gamma-1}\bigg(\frac{dp}{d\theta}-\gamma\frac{p}{\rho}\frac{d\rho}{d\theta}\bigg)
=-\frac{3f}{2}v_\phi T_{r\phi},
\end{equation}
where $r\phi$-component of stress tensor is: 
\begin{equation}
 T_{r\phi}=-\frac{3v_\phi}{2v_K}\alpha (p+\frac{B^2}{8\pi})
\end{equation}

The Faraday's induction equation with replacing radial dependency of quantities from Eq. (5), will give us
\begin{equation}
v_\theta B_r-v_rB_\theta=\frac{\eta}{r}(\frac{n-1}{2}B_\theta+\frac{dB_r}{d\theta}),
\end{equation}
\begin{displaymath}
\frac{d}{d\theta}\bigg[( v_\phi B_\theta - v_\theta B_\phi )
+\frac{\eta}{r\sin\theta}\frac{d}{d\theta}(B_\phi \sin\theta )\bigg]
\end{displaymath}
\begin{equation}
=\frac{n}{2}\bigg[v_\phi B_r-(\frac{n-1}{2}\frac{\eta}{r}+ v_r )B_\phi\bigg].
\end{equation}
Divergency-free equation, $\nabla\cdot\textbf{B}=0$, with applying Eq. (5) becomes,
\begin{equation}
 B_r(\theta)=\frac{1}{\sin\theta}\frac{d\psi(\theta)}{d\theta}, \hspace*{0.25cm}B_\theta(\theta)=\frac{n-3}{2\sin\theta}\psi(\theta).
\end{equation}
The last relation that is necessary to complete the set of equations, is to define the magnetic diffusivity, $\eta$. It is usually both the kinematic viscosity $\nu$ and the magnetic diffusivity $\eta$ are determined by a turbulence in accretion discs. Therefore, it is physically reasonable to write a similar formula for $\eta$  such as $\nu$ via the $\alpha$-prescription of Shakura \& Sunyaev (1973) as follows (Bisnovatyi-Kogan \& Ruzmaikin 1976) 
\begin{equation}
\eta(r,\theta)=\eta_0 \frac{r}{\rho v_K}(p+\frac{B^2}{8\pi}),
\end{equation}
where the dimensionless coefficient $\eta_0$ is assumed to be constant in the flow.

 In the next section, we will discuss boundary conditions to solve these differential equations numerically to obtain  $\rho(\theta), v_r(\theta), v_\theta(\theta), v_\phi(\theta), p(\theta), \psi(\theta)$ and $B_\phi(\theta)$. 

\section{Boundary Conditions} 
The relations of (6)-(15) provide a set of ODEs which needs to have boundary values for all quantities. It is very common to use reflection symmetry about the equatorial plane ($\theta=90^\circ$), which provides,
 \begin{equation}
 v_{\theta}=\frac{dv_r}{d\theta}=\frac{dv_\phi}{d\theta}=\frac{d\rho}{d\theta}=\frac{dp}{d\theta}=0,
 \end{equation}
 although we don't expect flow to be the same in both sides of the mid-plane, we just suppose the flow is locally symmetric about the equatorail plane. We should also define boundary value of magnetic field components. 
\begin{figure*}
\centering
\includegraphics[width=180mm]{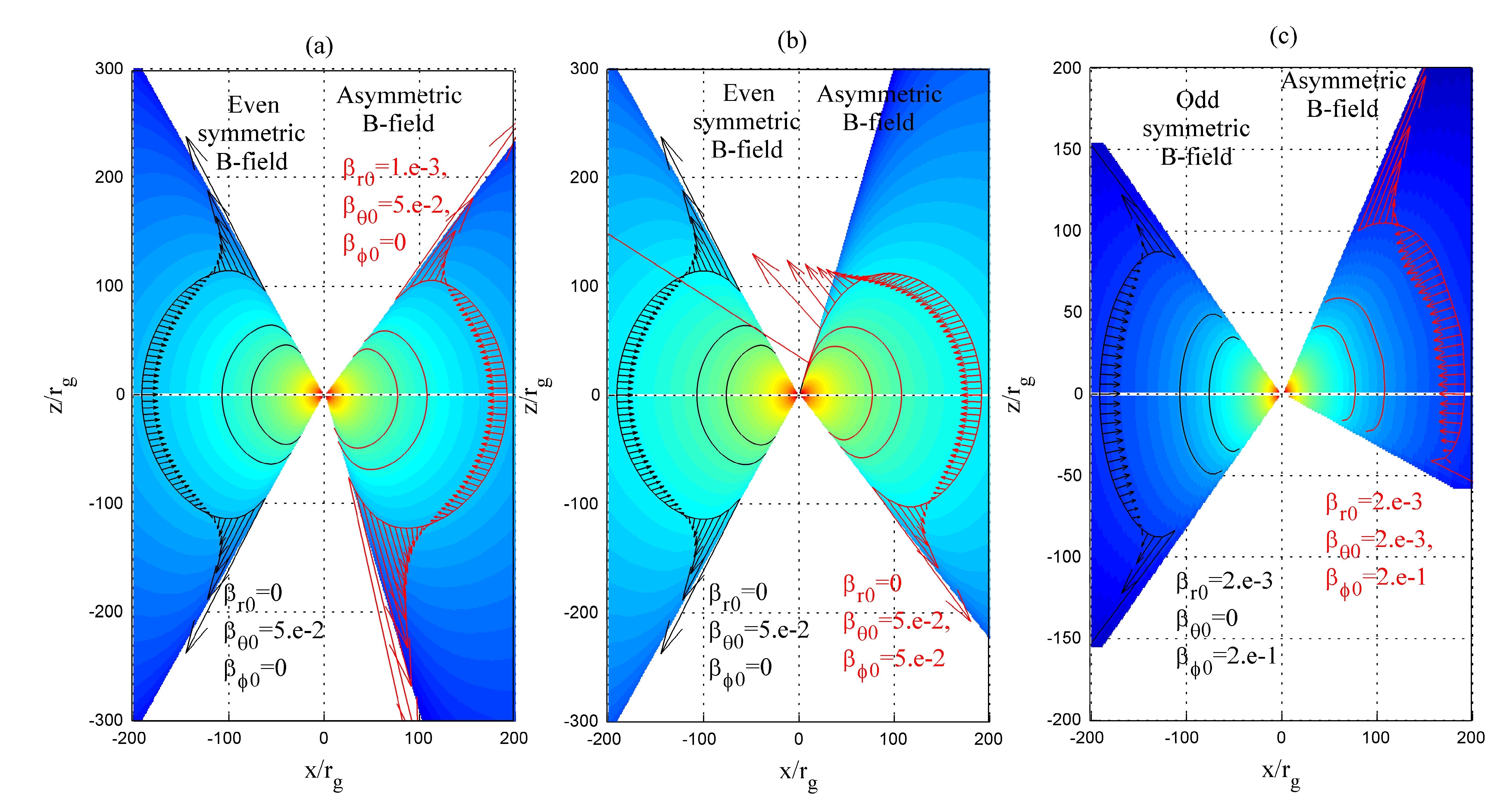}
 \caption{Density contours (3 black lines) and velocity field (arrows) $\textbf{v}=(v_r,v_\theta)$ in the meridional plane,  logartithm of density (colour), for a typical solution with  $n=1.2, f=1, \gamma=\frac{5}{3}, \alpha=\eta_0=0.1$. The left-hand panels show symmetrical magnetic fields. }\end{figure*}
 
To find proper boundary conditions for the magnetic field components, we firstly refer to the fashionable symmetric configuration proposed by Lovelace, Wang \& Sulkanen (1987)  and Wang, Sulkanen \& Lovelace  (1990).  According to their works, the magnetic field can have odd or even symmetry with respect to the equatorial plane. The case of odd field symmetry corresponds to the odd magnetic flux function. We have shown these configurations in figures 1-2. According to these figures, we see in an even symmetric field, $B_r$ and $B_\phi$ are in opposite direction above and below the equator. So they must become zero passing through the mid-plane. On the other hand, these two components don't change their signs in odd symmetry and just $B_\theta$ becomes zero at the equator and changes its sign.
 
 In the third case, we can imagine an asymmetric field which is superposing of two different fields with odd and even symmetric configurations. In this general situation, there are not any restrictions on the field's components at $z=0$ or $\theta=\pi/2$, so all of them might be nonzero there. However, we should know their values at the mid-plane due to boundary conditions. To do that, we can use the magnetic pressure conception ($p_m=B^2/(8\pi)$). It is common to compare magnetic pressure with gas pressure  $p_m/p_g=\beta$\footnote{Note that the standard definition of plasma parameter is usually $\beta=p_{gas}/p_{mag}$}. Similar to this definition we can present the  $\textbf{B}$components in the equator as,
 \begin{equation}
 B^2_{r0}=8\pi\beta_{r0} p_0, \hspace*{0.5cm}B^2_{\theta0}=8\pi\beta_{\theta0} p_0, \hspace*{0.5cm} B^2_{\phi0}=8\pi\beta_{\phi0} p_0,
 \end{equation}
where, zero index shows the quantity belongs to the equator. In the next step, we calculate the derivatives of $B_r, B_\theta$ and $B_\phi$ at $\theta=90^\circ$ by substituting boundary conditions of eqn.(16), (17) in basic equations (6)-(15),
\begin{equation}
\frac{dB_r}{d\theta}|_0=-\bigg[\frac{n-1}{2}+\frac{r\rho_0v_{r0}}{\eta_0 p_0(1+\beta_0)}\bigg]B_{\theta0},
\end{equation}
\begin{equation}
  \frac{dB_\theta}{d\theta}|_0=\frac{2}{n-3}B_{r0},
\end{equation}
  \begin{equation}
 \frac{dB_\phi }{d\theta}|_0=\frac{r\rho_0 v_{r0}}{\eta_0 p_0(1+\beta_0)}\frac{B_{r0}B_{\theta0}}{B_{\phi0}},
 \end{equation}
 where $\beta_0=\beta_{r0}+\beta_{\theta0}+\beta_{\phi0}$. 
 As it was mentioned in section 2, for zero divergency of the magnetic field we should use the flux function, $\psi$, in our numerical solution and we need to know this function in the mid-plane. We can easily determine the value of magnetic flux and its first and second derivatives at the equator according to Eq. (14) and using the relations above,
 \begin{displaymath}
 \psi_0=\frac{2}{n-3}B_{\theta 0}=\frac{2}{n-3}\sqrt{8\pi\beta_{\theta0} p_0}
 \end{displaymath}
 \begin{displaymath}
\frac{d\psi}{d\theta}|_{90^\circ}=B_{r0}=-\sqrt{8\pi\beta_{r0} p_0}
 \end{displaymath}
 \begin{displaymath}
 \frac{d^2\psi}{d\theta^2}|_{90^\circ}=-\bigg[\frac{n-1}{2}+\frac{r\rho_0v_{r0}}{\eta_0 p_0(1+\beta_0)}\bigg]\sqrt{8\pi\beta_{\theta0} p_0}
 \end{displaymath}
  
 We can choose $\rho_0=1$ for imposing a characteristic scale density at $\theta=90^\circ$  (Xu \& Wang 2007). Then $v_{r0}, v_{\phi0}$ and $p_0$ can be found by using the above results in Eq.(7), (9) and (10). In the following, we will try to examine different values of $\beta_{i0}$ to see how the vertical structure will change in the both sides of equatorial plane. 
\begin{figure*}
\centering
\includegraphics[width=180mm]{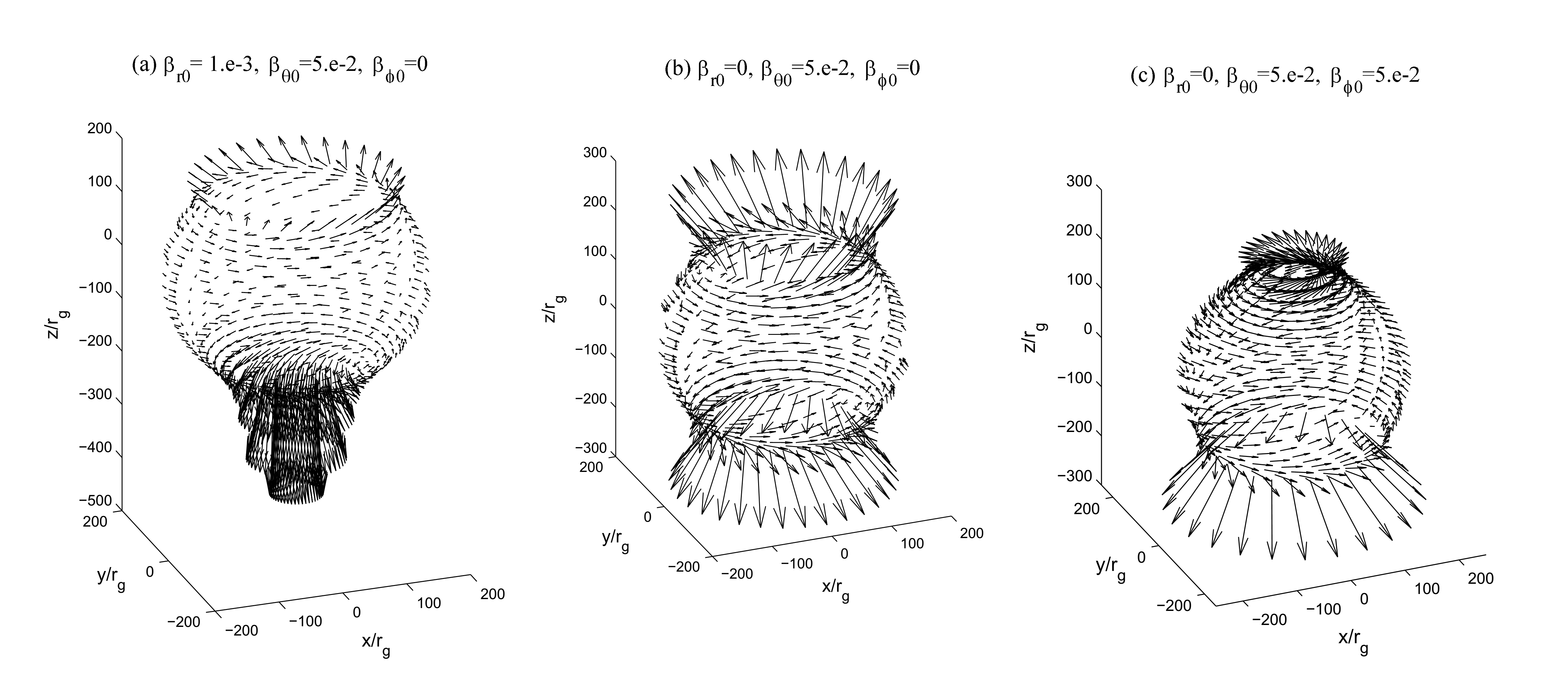}
 \caption{3D presentation of velocity field in the presence of three different magnetic fields: a) asymmetric field with $\beta_{\phi0}=0$ b) symmetric field with $\beta_{r0}=\beta_{\phi0}=0$ c) asymmetric field with $\beta_{r0}=0$. Other parameters are $n=1.2, f=1, \gamma=\frac{5}{3}, \alpha=\eta_0=0.1$.}
\end{figure*}
\section{Results}
With the complete set of equations (6)-(15) and proper boundary conditions, we can numerically integrate all equations for unknowns quantities in the vertical direction. This integration has been done firstly over the top side of the disc and secondly over the bottom side, starting from the mid-plane ($\theta=90^\circ$). It should be mentioned the integrations would be stopped by a numerical error because the  density or gas pressure becomes zero in a certain angle called opening angle, $\theta_b$ (Xue \& Wang 2005; Jiao \& Wu 2011). 
Moreover, we find the flow is separated to two regions between $\theta=90^\circ$ and $\theta=\theta_b$ (and between $\theta=90^\circ$ and $\theta=\pi-\theta_b$ bottom side of the disc). The separator is at a certain latitude ($\theta=\theta_0$ top side and $\theta=\pi-\theta_0$ bottom side) where radial velocity is zero. Upside and downside the disc near the equatorial plane, the flow is accreting to the centre $v_r<0$, so we call it inflow. Outside the inflow region, from $\theta_0$ to $\theta_b$ the flow is in the opposite direction and moves outward. Hence the outflow is formed there. In this work, we notice that the sides of the disc are not similar. So each side has different thickness with a bit different dynamics as seen in the following figures.

In the solution presented by Naryan \& Yi (1994), it was assumed that the mass accretion rate is constant with radius. Consequently the density follows a power-law function of radius with a constant index, $\rho\propto r^{-n}$, $n=1.5$. However, several hydrodynamics and magnetohydrodynamics simulations clearly pointed out that accretion rate decreases with decreasing radius, with density index, $n\simeq 0.5-1$ (Stone, Pringle \& Begelman 1999; Igumenshchev \& Abramowicz 1999, 2000; Stone \& Pringle 2001; Machida, Matsumoto \& Mineshige 2001; Yuan et al. 2012a,b). In adiabatic inflow/outflow solutions (ADIOS: Blandford \& Begelman 1999, 2004), this varying inflow rate is due to a genuine mass loss in a wind. Recent improvements of ADIOS (Begelman 2012) have shown that n should be roughly close to unity. Following SA16, two different values for density index have been adopted, i.e. $n=0.85$ and $n=1.2$.  SA16 have shown that  the stronger outflows arise from discs with larger density index  $n=1.2$ (see figure 1 in SA16).


\begin{figure*}
\centering
\includegraphics[width=180mm]{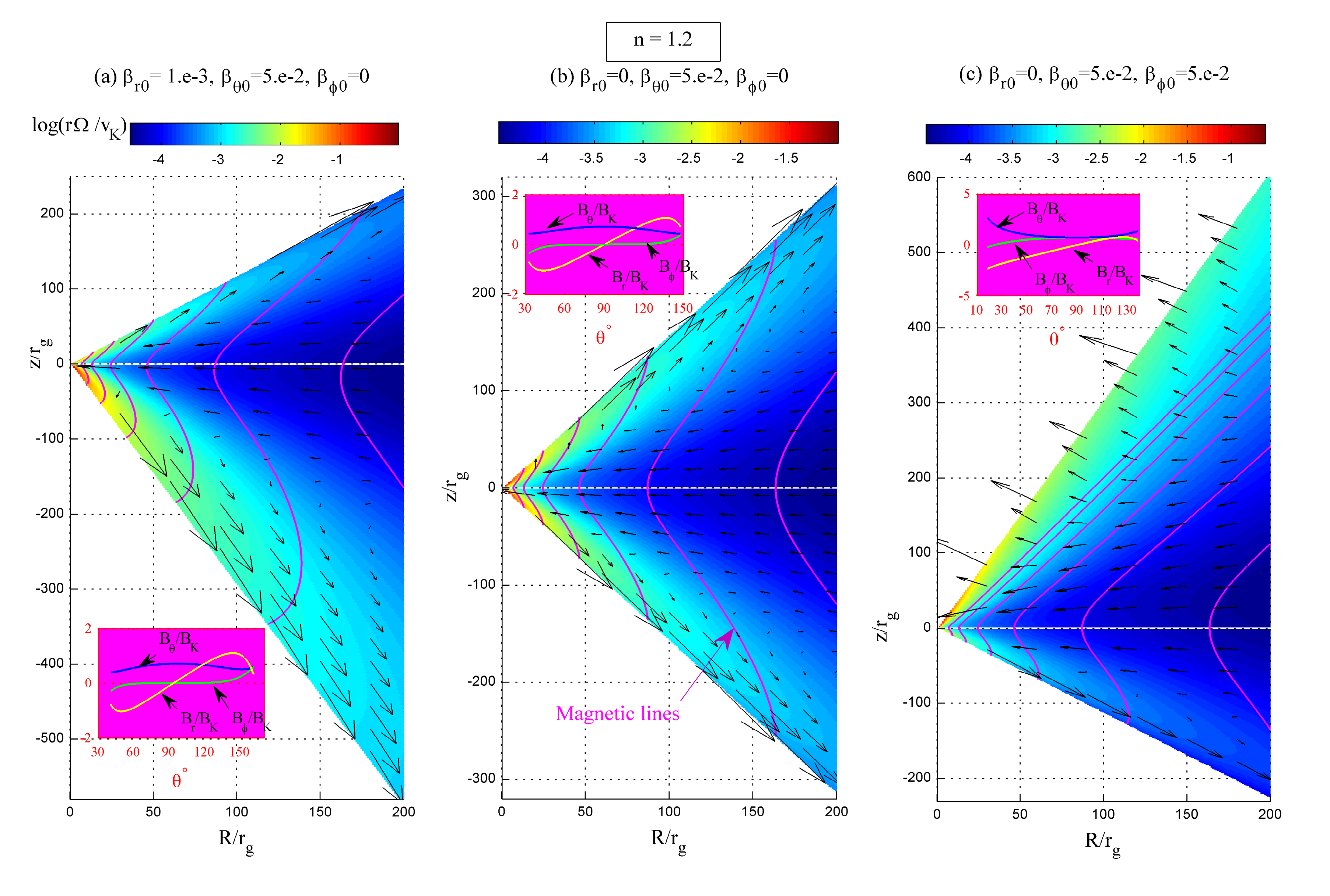}
 \caption{Velocity field (black arrows), 2D presentation of magnetic field lines (pink lines), logarithm of angular velocity $\Omega(\theta)=v_\phi(\theta)/\sin\theta$ (colour) with  $n=1.2, f=1, \gamma=\frac{5}{3}, \alpha=\eta_0=0.1$. The small pink panels show all components of magnetic field with respect to polar angle. $v_K$ is Keplerian velocity and $B_k$ means $\sqrt{\rho_0 v_K^2}$.  }
\end{figure*} 
In order to study asymmetric configuration for magnetic field, we consider the magnetic parameters of $\beta_{r0}, \beta_{\theta0}$ and $\beta_{\phi0}$ with similar ones in symmetric fields but with small deviations. As seen there are 3 panels in figure 3, that each panel consist of 2 magnetic fields (in the left side, upside $z>0$ and downside $z<0$ are the same but in the right side the disc has different heights). The left side of each panel displays the vertical structure of disc in the presence of a symmetric field while the left side of the panel represents the asymmetric disc due to an asymmetric magnetic field. We have examined 3 possible situations of asymmetric configuration of magnetic field. In a and b panels of Fig.3, the asymmetric fields provided by small deviations from an even symmetry B-field with a) $\beta_{r0}=1.e^{-3}$, $\beta_{\phi0}=0$ and b)$\beta_{r0}=0$, $\beta_{\phi0}=5.e^{-2}$. As a matter of fact, $B_{r}$ and $B_{\phi}$ are zero at the equator in the corresponding even symmetry field with $\beta_{r0}=\beta_{\phi0}=0$ and $\beta_{\theta0}=5e^{-2}$ which is presented in the right sides of these two panels. For this even field, our calculations end at equal opening-angle above and below the equator, but when radial component (on the right-hand side of Fig3.a) or toroidal one (on the right-hand side of Fig3.b) are not zero at $\theta=90^\circ$, they make disc's behaviour change particularly near the outflow regions. In the middle panel, $B_{\phi0}$ is nonzero and makes the disc expand in the northern hemisphere and inversely compress in the southern one. In the upper side of disc, we see matter ejects in the meridional direction rather than radial one.  In Fig.3 (c), we have compared an odd symmetric magnetic field with an asymmetric one.

 It is interesting to point out that Fiege \& Henriksen 1996a, b have studied bipolar outflows from protostars based on the idea of steady quadrupolar circulation and obtained the general properties for the outflow region similar to SA16' result in the presence of odd symmetric magnetic field.
 
  SA16 have proposed a constrain for odd symmetry case in order to have strong outflow: $B_\phi$ must be dominant component. Using this condition and setting $\beta_{\theta0}=\beta_{r0}$ we have investigated another asymmetrical configuration for the magnetic field. In Fig.3 (c), it's clear that adding the third part of B-field makes the disc compress from one side but at the same time it stretches another side of the disc. Furthermore, the outflow region doesn't differ significantly because $\beta_{\theta0}$ isn't large enough to cause huge changes. The colours in this figure display logarithm of density. We have shown the 2D velocity fields on the isodensity line. 
\begin{figure*}
\centering
\includegraphics[width=180mm]{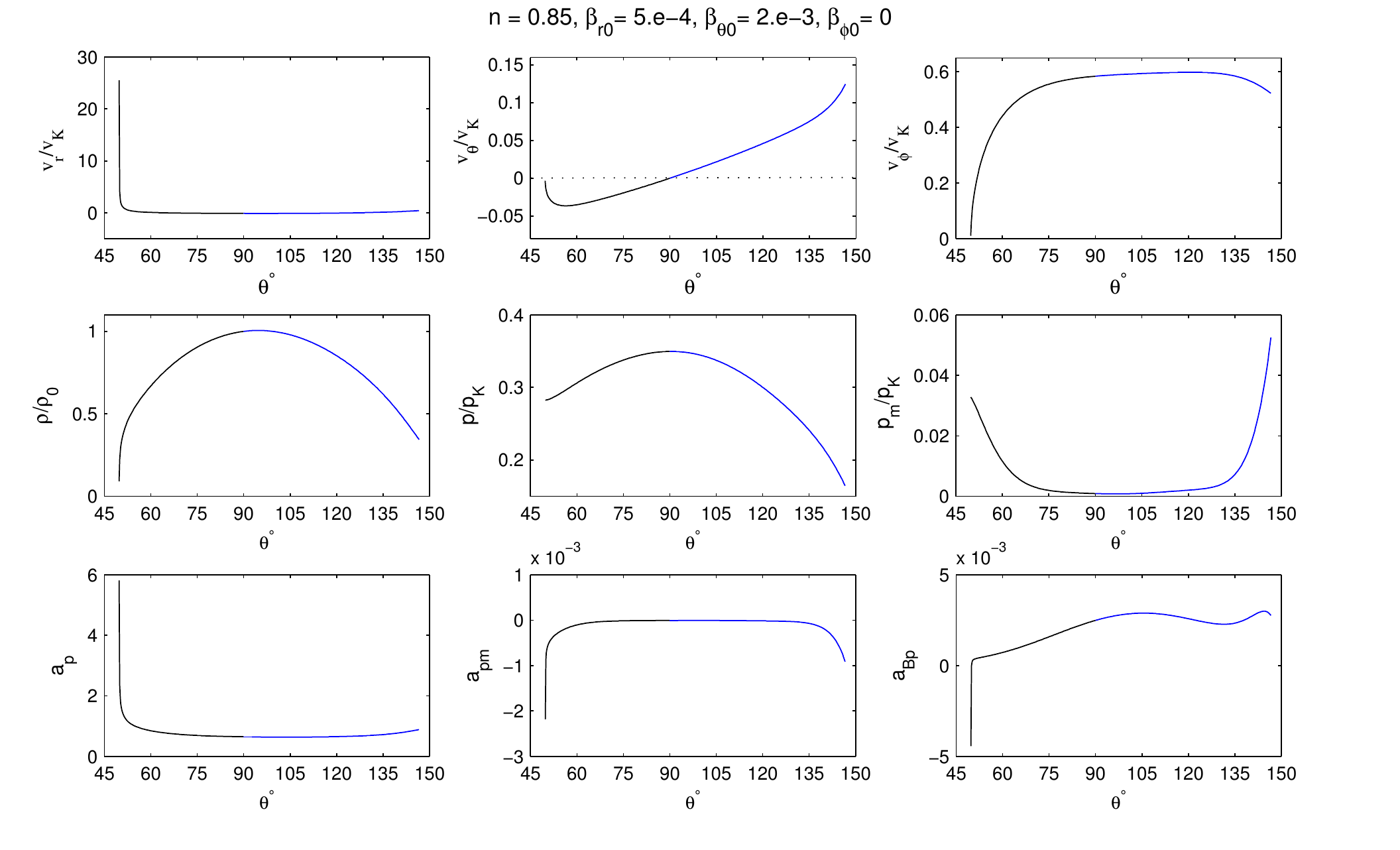}
 \caption{Self-similar solutions corresponding to $n=0.85, \gamma=5/3, \alpha=\eta_0=1$ and $f=1$. The magnetic parameters  provide an asymmetric field which is based on a small deviation from a primary even symmetry configuration ($\beta_{r0}\neq0$). The panels in the third row show the radial accelerations including $a_p=(n+1)p/\rho$, $(n-1)p_m/(8\pi\rho)$, $a_{Bp}=[2 B_{\theta} dB_r/d\theta+(1-n)B_r^2]/(8\pi\rho)$ in Eq.(23). Here $v_K$ is Keplerian velocity and the fiducial pressure is defined by $p_K=\rho_0 v_K^2$. Notice in this figure, we have ignored the constraint on Mach number.   }
\end{figure*} 
\begin{figure*}
\centering
\includegraphics[width=180mm]{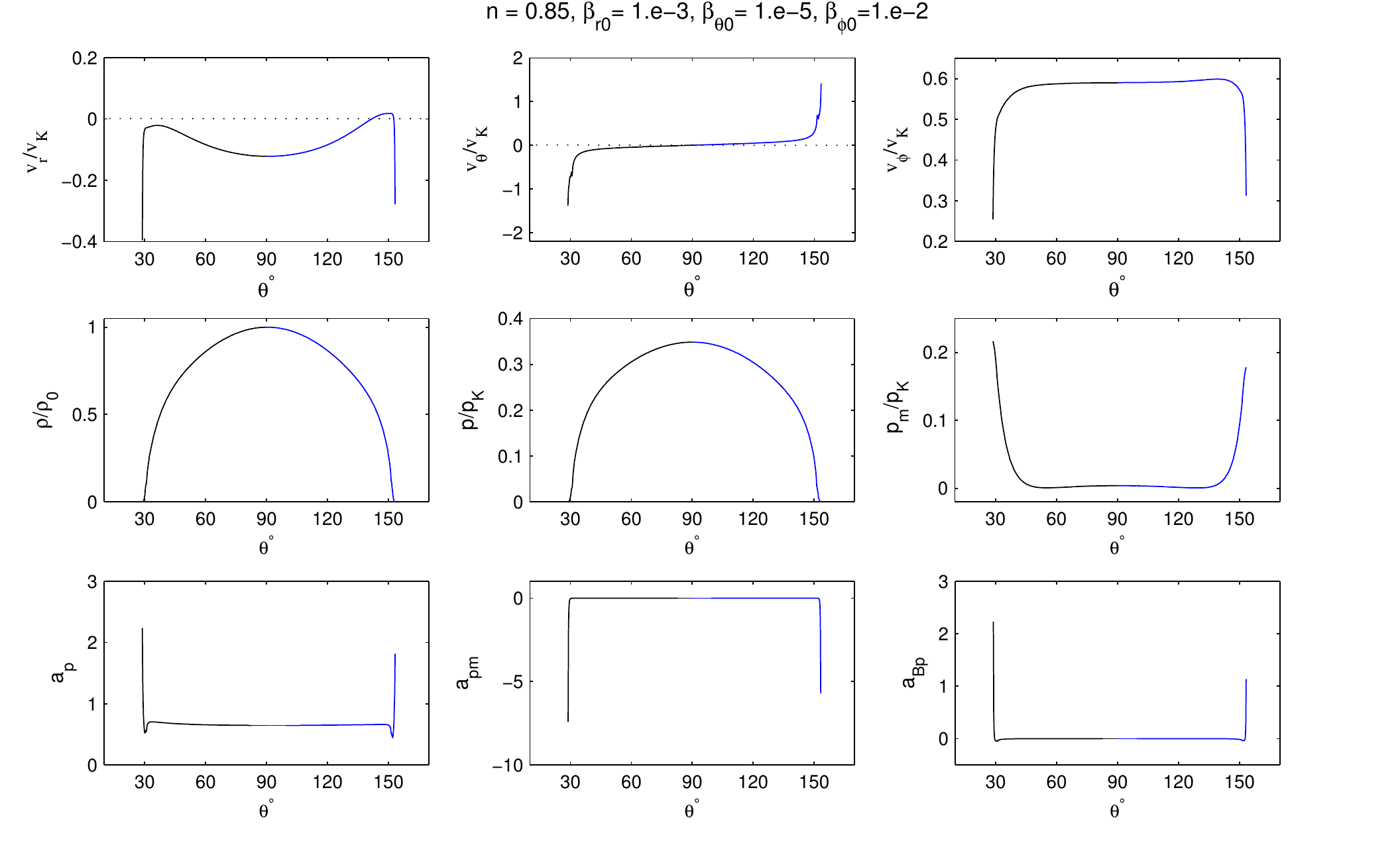}
 \caption{Self-similar solutions corresponding to $n=0.85, \gamma=5/3, \alpha=\eta_0=1$ and $f=1$. The magnetic parameters  provide an asymmetric field which is based on a small deviation from a primary odd symmetric field ($\beta_{\theta0}\neq0$). The panels in the third row show the radial accelerations including $a_p=(n+1)p/\rho$, $(n-1)p_m/(8\pi\rho)$, $a_{Bp}=[2 B_\theta dB_r/d\theta+(1-n)B_r^2]/(8\pi\rho)$ in Eq.(23). Here $v_K$ is Keplerian velocity and the fiducial pressure is defined by $p_K=\rho_0 v_K^2$. Notice in this figure, we have ignored the constraint on Mach number.   }
\end{figure*} 

Figure 4 shows 3D velocity fields for two deviations from the even symmetric magnetic field which is defined by  $\beta_{r0}=\beta_{\phi0}=0$ and $\beta_{\theta0}=5\times 10^{-2}$ (the middle plot of Fig. 4). 
 
 Figure 5 represents 2D view of magnetic field lines in $x-z$ plane on the colour background. The colours show logarithm of angular velocity $\Omega(\theta)$. The small pink image inside each panel displays vertical trends of all components of \textbf{B}. To have a better judgment, the velocity field is plotted in this figure in the regular intervals. If we move along these 2D magnetic lines, angular velocity is almost constant, hence $B_r$ and $B_\phi$ have the same sign (see pink diagrams inside panels just in Fig5.(a) $B_r$ and $B_\phi$ have different signs in some parts above the mid-plane). Thus the toroidal component of magnetic field is not due to the shear of the accretion flow in all parts of the disc. Furthermore, in this figure it seems that Ferraro's theorem is correctly playing a role in keeping the the rotational velocity constant on a field line in a steady state. 
  The other feature of this figure is that the stronger outflows appear on the brighter colours that means $\Omega$ or centrifugal force plays an important role to drive outflows.

Following Yuan et al.'s simulations (2012a,b), in the rest figures we have adopted $n=0.85$. To have solution containing outflows, this selection of density index restricts us to select the 3 magnetic parameters to be very small, i.e. $\beta_{i0} <10^{-2}$ (SA16). To explain more, we consider the radial component of the momentum equation,
 \begin{displaymath}
v_\theta\frac{dv_r}{d\theta}=\frac{1}{2}v^2_r+v_\theta^2- v^2_k+v^2_\phi+(n+1)\frac{p}{\rho}\end{displaymath}
  \begin{equation}
  \hspace*{1.5cm}  +\frac{n-1}{8\pi}\frac{p_m}{\rho}+\frac{1}{8\pi\rho}\bigg[2 B_\theta\frac{dB_r}{d\theta}+(1-n)B_r^2\bigg],
\end{equation}
 where $p_m$ is the magnetic pressure. We know in this equation, $v_\phi^2/r$ is the centrifugal acceleration per unit mass and the term including $p$ are always positive factors for driving outflow. On the other hand, the magnetic terms in this relation do not have a constant effect on the outflow region. As we see the term including $p_m$ has the factor of $n-1$, so this term helps to make $v_r$ positive if the density index is supposed to be greater than unity. For instance, in the presence of a purely toroidal magnetic field, Fig. (1) of Mosallanezhad et al. (2014) with $n=0.85$ implies the stronger outflows appear in weaker $B_\phi$ and also Fig.(3) of them shows $v_r$ is more positive in the outer region when $n-1$ is smaller (or less negative). Therefore, the magnetic pressure is a negative term for $v_r$ in the presence of a purely toroidal field with $n<1$. Now for adopting $n<1$, we know the magnetic parameters should be selected very small in order to reach positive radial velocities in the outer parts of the disc. Nevertheless, for the general situation it is not easy to determine the effect of the magnetic field because of the last term in Eq.(21). Notice, the factor of $(n-1)/8\pi$ is very small, almost equal to 0.005 with $n=0.85$, hence $p_m$ might have noticable effect where $\rho\rightarrow 0$ just near the surface of the disc. 
 

 The $\theta-$dependency of velocity components, density, gas and magnetic pressures are presented in Fig.6 and 7 for 2 different sets of magnetic parameters. Outstanding feature of Fig.6 is very large value of radial velocity at $\theta=47^\circ$ which is supersonic. Fiege \& Henriksen (1996b) by using quadrupolar symmetry have had similar opening angle for protostellar bipolar outflows. The divergently large radial outflow at small angle is also found in their work. In Fig.6, the other components of velocity are almost zero in that angle. This situation demonstrates a very strong outflow in the top of the disc which moves away from the centre with the whole kinitic energy of the flow. However, the outflow on the other side of the disc is very weak. This huge differences between two sides of the disc can be applied to explain one-sided jets from FR II (Wang et al. 1992).    In another study about outflows, Lery et al. (1999) found very similar profile for radial velocity, although they had ignored viscosity and assumed radiative cooling to be dominant. 


In the middle row panels of  Fig.6, the density and gas pressure decrease from the mid-plane towards the surface, whereas the magnetic pressure has opposite trend and becomes larger in the outflow region (the same result is achieved by Samadi et al. 2014). 
In addition, the peaking pressure at intermediate angles in fig.6 is consistent with compressing the disc.
In the last row of Fig.6, we have plotted 3 terms in Eq.(21) versus the polar angle. We mentioned the main terms in Eq.(21) which affect the sign of $v_r$ are 1) centrifugal acceleration $v_\phi^2/r$, 2) thermal acceleration $a_p=(n+1)p/\rho$, 3) magnetic pressure term $a_{pm}=(n-1)p_m/(8\pi\rho)$, 4) polar magnetic term  $a_{Bp}=[2 B_\theta dB_r/d\theta+(1-n)B_r^2]/(8\pi\rho)$. As seen in Fig.6, $a_p$ has the main role for driving outflow from both sides the disc. The rotational velocity is maximum in the midplane and dominant component of velocity at first though, it decreases outwards. So the centrifugal force has less effect to make $v_r$ positive in the outermost regions if we compare it with $a_p$. The magnetic terms $a_{pm}, a_{Bp}$ are 3 orders of magnitude less than thermal acceleration and can't directly change the slope of $v_r$ along $\theta-$direction. 
In this special case, magnetic field pressure has negative effect on emerging outflow but in some MHD outflow solutions (Lery et a. 1999, Henriksen \& Valls-Gabaud 1994, Fiege \& Henriksen 1996a, b)
 magnetic pressure becomes the main force which drives outflows. Lery et al. (1999) found a pressure driven analytic solution just as we have found in Eq. (21) to be the driving forces. 
Figure 7 shows that the negative acceleration of magnetic pressure in the total radial acceleration that causes $v_r$ end up negative in the outflow region. In this case, the magnetic pressure becomes large enough near the disc's surface and acts in the opposite direction of gas pressure. On the other hand, the poloidal components of magnetic field produce a positive large force but not as large as the force of magnetic pressure. Consequently, in this case, the total effect of the magnetic terms is not positive on the surface. 
       
In figures 8-14, we stop our calculations where $M=v_{total}/c_s\geq1$. In Fig.8, we have examined the variation of velocity field in the presence of different asymmetric fields. The black curve in panels (a),(b) show an even symmetric field corresponding to $\beta_{r0}=\beta_{\phi0}=0$ and $\beta_{\theta0}=2\times 10^{-3}$. According to the panel (a), with nonzero radial component of the magnetic field at $\theta=90^\circ$, the radial velocity does not have the reflection symmetry about the equator and significant changes appear with larger value of $\beta_{r0}$. In the panel (b), another small change in the input parameters gives different asymmetric disc. Although $v_r$ is positive above and below the disc midplane, the half-thickness of each side is not the same as the other side of the disc. In the panel (c), the black line presents an odd symmetric field corresponding to $\beta_{\theta0}=0$, $\beta_{r0}=1.e^{-3}$ and $\beta_{\phi0}=1.e^{-2}$. In this case, we face to a solution without outflow. The variation of $\beta_{\theta0}$ yields new solution with asymmetric outflow from one side or two sides of the disc.    

According to figures 3 and 8, we see the half-thickness of the disc is different in each side for both $n>1$ and $n<1$. We realise one side is thicker than the other side but what determines the disc's thickness? Comparing Fig.3, 8 we find out when $\beta_{\phi0}$ is large enough (like 2 order of magnitude larger) in comparison with $\beta_{r0}$ and $\beta_{\theta 0}$ it will make disc thicker in the upper side and thinner in the other side. If $\beta_{\phi0}$ isn't large enough then we can expect the disc's half-thickness to be the same or become thinner above the mid-plane if $\beta_{\phi0}\sim\beta_{r0}$ and $\beta_{\theta0}$ isn't too small.  

\begin{figure*}
\centering
\includegraphics[width=180mm]{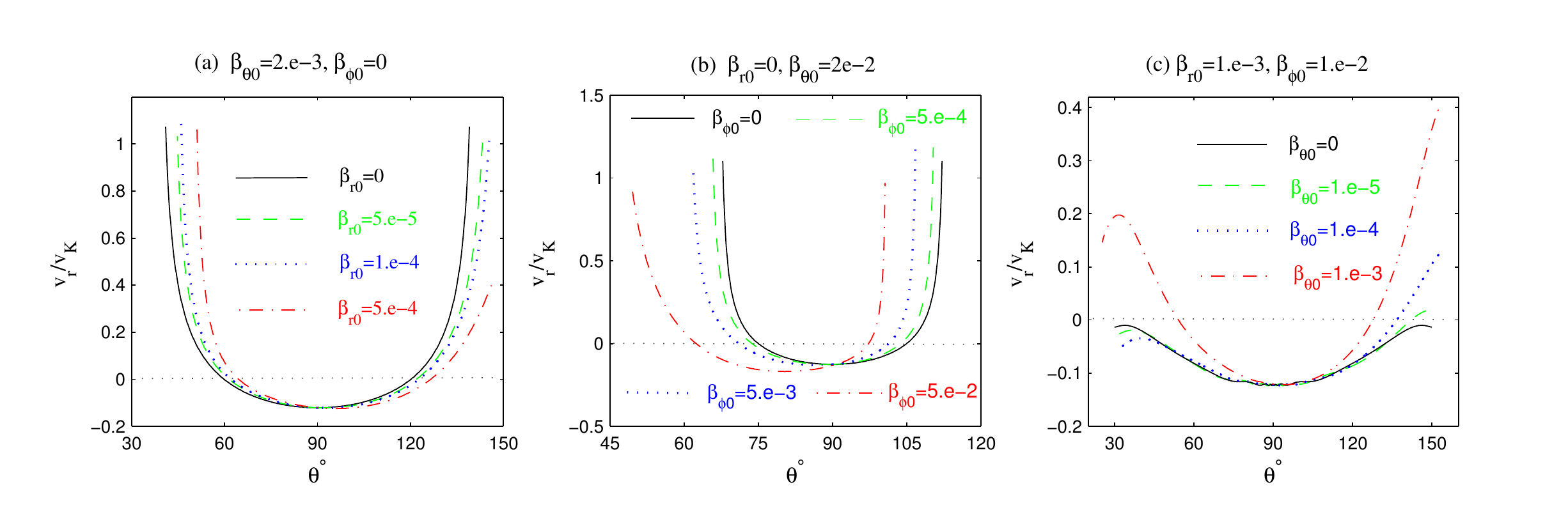}
 \caption{Distribution of radial ($v_r$), meridional ($v_\theta$) and rotational ($v_\phi$)  velocities.  Here $n=0.85, \gamma=5/3, \alpha=\eta_0=1$ and $f=1$. $v_K$ is Keplerian velocity. In the left panel we have examined deviations from even symmetry with fixing $\beta_{\phi0}=0$, the middle panel shows another deviation from even $\textbf{B}$- field with constant $\beta_{r0}=0$ and the right panel is plotted with a deviation from odd symmetric case.  }
\end{figure*} 
\begin{figure*}
\centering
\includegraphics[width=180mm]{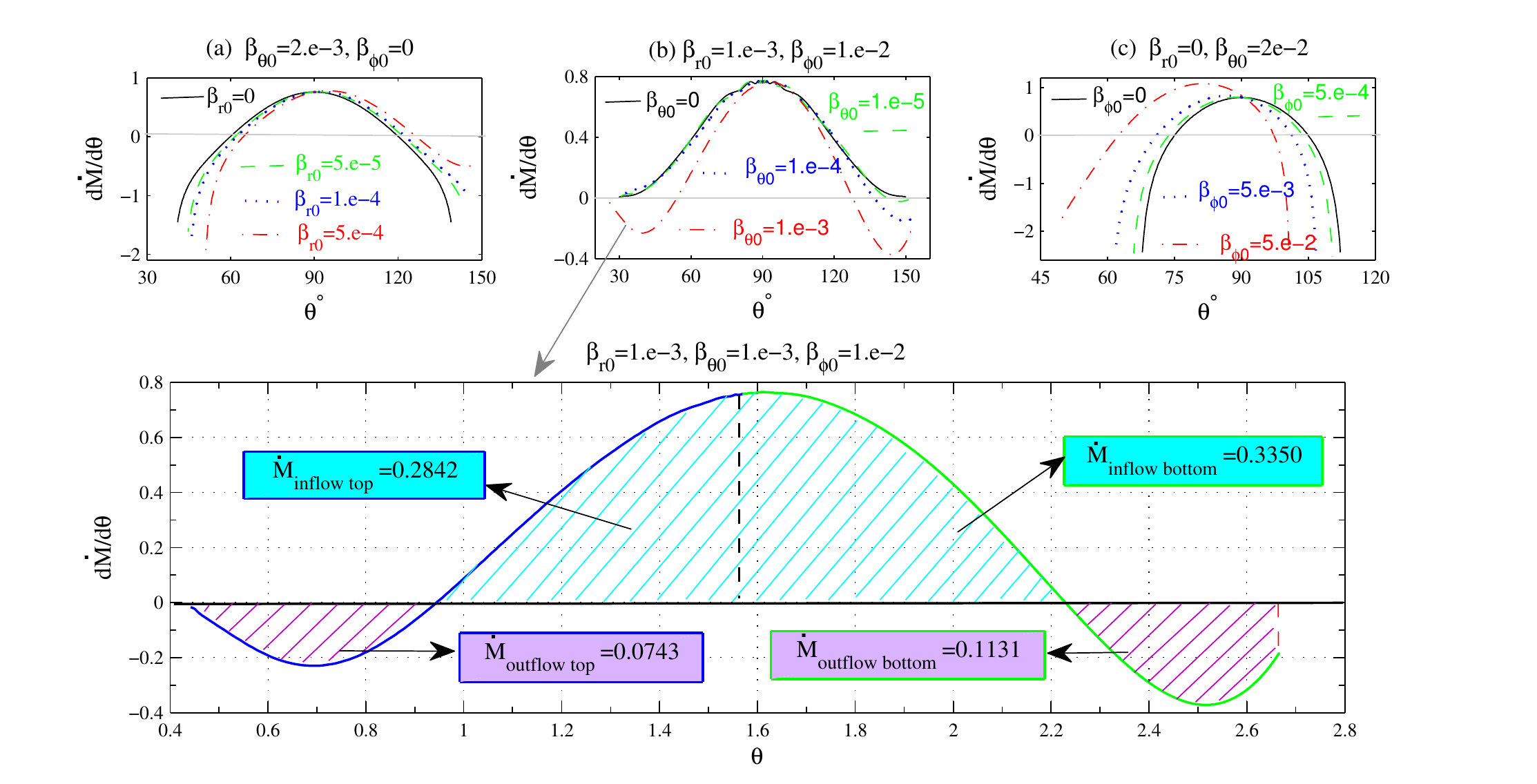}
 \caption{The distribution of mass rate differential along $\theta-$direction in the presence of an asymmetric magnetic field. Here $n=0.85, \gamma=5/3, \alpha=\eta_0=1$ and $f=1$. The area above the horizontal axis shows total accretion rate whereas the two bottom areas are negative and belongs to outflow region. The horizontal axis is polar angle specified in radian. }
\end{figure*}

\subsection{Mass Accretion Rate}

 In hot accretion flows, we know the accretion rate varies with respect to the radius. One reason for this change in $\dot{M}$ is due to outflow (Blandford \& Begelman 1999, 2004). Using self-similar solutions, we calculate this quantity as
 \begin{equation}
 \dot{M}=-2\pi{r^{1.5-n} }\int \rho(\theta) v_r(\theta)\sin\theta  d\theta.
 \end{equation} 
 Then we obtain,
 \begin{equation}
d\dot{M}/d\theta=-2\pi \rho(\theta) v_r(\theta)\sin\theta, 
 \end{equation}
  where $d\dot{M}/d\theta$ is dimensionless. Figure 9 displays the distribution of $d\dot{M}/d\theta$ along the $\theta-$direction. 
  It's clear that $v_r$ is negative from the midplane and somewhere near the surface it becomes zero and then increases to positive values. In addition, $\rho$ is maximum at $\theta=90^\circ$, hence $d\dot{M}/d\theta|_{\theta=90}$ is positive and maximum. In the bottom panel of  Fig.9, we have shown the total mass inflow/outflow rates with respect to polar angle (in unit of radian) in every region inside a disc by calculating the area bounded between the graph of $d\dot{M}/d\theta$ and $\theta$ axis.  This figure shows that in both sides of the disc a noticeable amount of mass is carried out by outflow (Above the equator, the outflow rate is almost a quarter of inflow rate and in the bottom side of the disc, it is a bit larger, almost a third of inflow rate). 
\subsection{Magnetic Field Configuration}
To study the magnetic field configuration within the disc we will
look at the magnetic field lines, which satisfy the following equation: 
\begin{equation}
\frac{dr}{B_r}=\frac{rd\theta}{B_\theta}=\frac{r\sin\theta d\phi}{B_\phi},
\end{equation}

 In Fig. 10-12, we have shown the magnetic field lines related to Fig. 8. An asymmetric magnetic field is due to nonzero $\beta_{r0}$ while $\beta_{\phi0}=0$ in Fig 10. We can see lines orbit clockwise in the northern hemisphere and counterclockwise in the southern hemisphere (see the middle panels which shows $x-y$ plane). It means $B_\phi$ is negative in the upper side and negative in the other side which is the feature of even symmetry. For very small $\beta_{r0}$ blue lines which belong to the northern hemisphere are a bit away from green lines which are displaying magnetic field of the southern hemisphere. In an even $\textbf{B}-$field, sign of $B_r$ changes throughout the equator. In the right panels, we can see $B_r$ has different directions in asymmetrical field with a small $\beta_{r0}$ too. However, with a big value of $\beta_{r0}$, the direction of lines changes twice in the southern hemisphere. This repetition of changing sign in one side of disc happens for $B_\phi$ too (according to the middle-bottom panel of Fig.10).  According to Eq. (22), wherever $B_r/B_\theta$ is positive $dr$ is positive too. If $B_\theta$ becomes zero in a certain $\theta$, it will yield an infinity radius. After that singularity $dr$'s direction changes so we can see green arrows which oriented downwards, convert to yellow arrows which direct to upwards in $x-z$ plane (the bottom-right panel of Fig. 8). We know $B_\theta$ becomes zero in a certain angle and it makes the direction of $d\phi$ change below the equator. Moreover, the bottom panels let us think the configuration is converting to an odd symmetry because $B_\phi$ and $B_r$ have the same direction in an odd symmetry. When $\beta_{\theta0}<\beta_{\phi0}$ the field is more likely an odd symmetry.
Other information from this figure is that $B_\phi$ is the dominant component of the field. 

Fig. 11 shows magnetic field lines with different asymmetric configuration which is created by nonzero $\beta_{\phi0}$. In the first two row panels we can see a very small deviation from even symmetry field, $\beta_{\phi0}$ is two and one order of magnitude less than $\beta_{\theta0}$. The lines in top and middle of this figure are almost vertical, it means $B_\theta$ is dominant not only in the boundary of equator but also every where inside the disc. The most deviation from even symmetry appear just in the bottom panels of this figure where $\beta_{\phi0}\sim \beta_{\theta0}$. We mentioned $B_r$ and $B_\phi$ are zero in the equatorial plane in even symmetry, and they changes their direction throughout $\theta=90$. Now, in the right panels which shows the picture of magnetic lines in $x-z$ plane. Blue lines say $B_r$ is directing towards the centre in $z>0$ and green lines show $B_r>0$ in $z<0$. But the bottom-right panel has an extra direction for $B_r$ in $z>0$. In this panel $B_r$ is in the same direction of $\hat{e}_r$ near the north pole then near the midplane, its direction changes.   In the top and bottom panels we can see  in $z>0$, $B_r<0$ (blue lines directing towards the centre in $x-z$ plane in the right panels) and $B_\phi<0$ (blue lines are clockwise in the middle panels) although they are very small. A huge difference is seen in the third row, with $\beta_{\phi0}=2.5\beta_{\theta0}$.  As we see in the $x-y$ plane, the blue magnetic line and green line are both counterclockwise (i.e. $B_\phi>0$). According to the bottom right panel, in $x-z$ plane, near $z=0$, we see $B_r<0$ in the both side of the disc. This case is roughly similar to an odd symmetry field whose $B_\phi$ and $B_r$ have constant direction both above and below the disc's mid-plane.  
\begin{figure*}
\centering
\includegraphics[width=170mm]{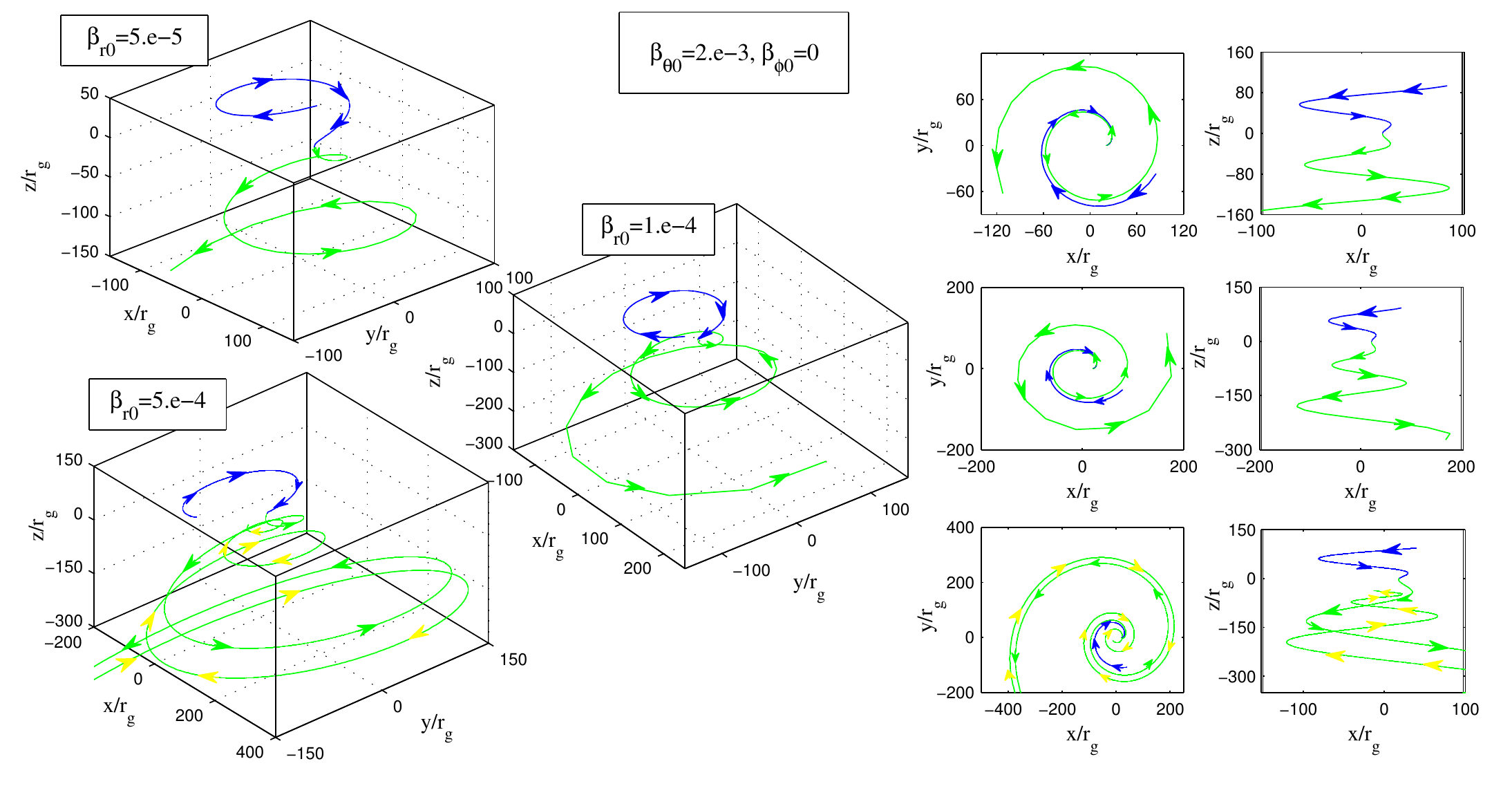}
 \caption{3D presentation of magnetic field lines for 3 different cases which display deviations from even symmetry where $n=0.85, \gamma=\frac{5}{3}, \alpha=\eta_0=0.1$ and $ f=1$. We have fixed $\beta_{\phi0}=0$ and $\beta_{\theta0}=2\times 10^{-3}$. The 2 columns panels in the right-hand side show the same magnetic lines in $x-y$ plane and in $x-z$ plane.  }
\end{figure*}
\begin{figure*}
\centering
\includegraphics[width=170mm]{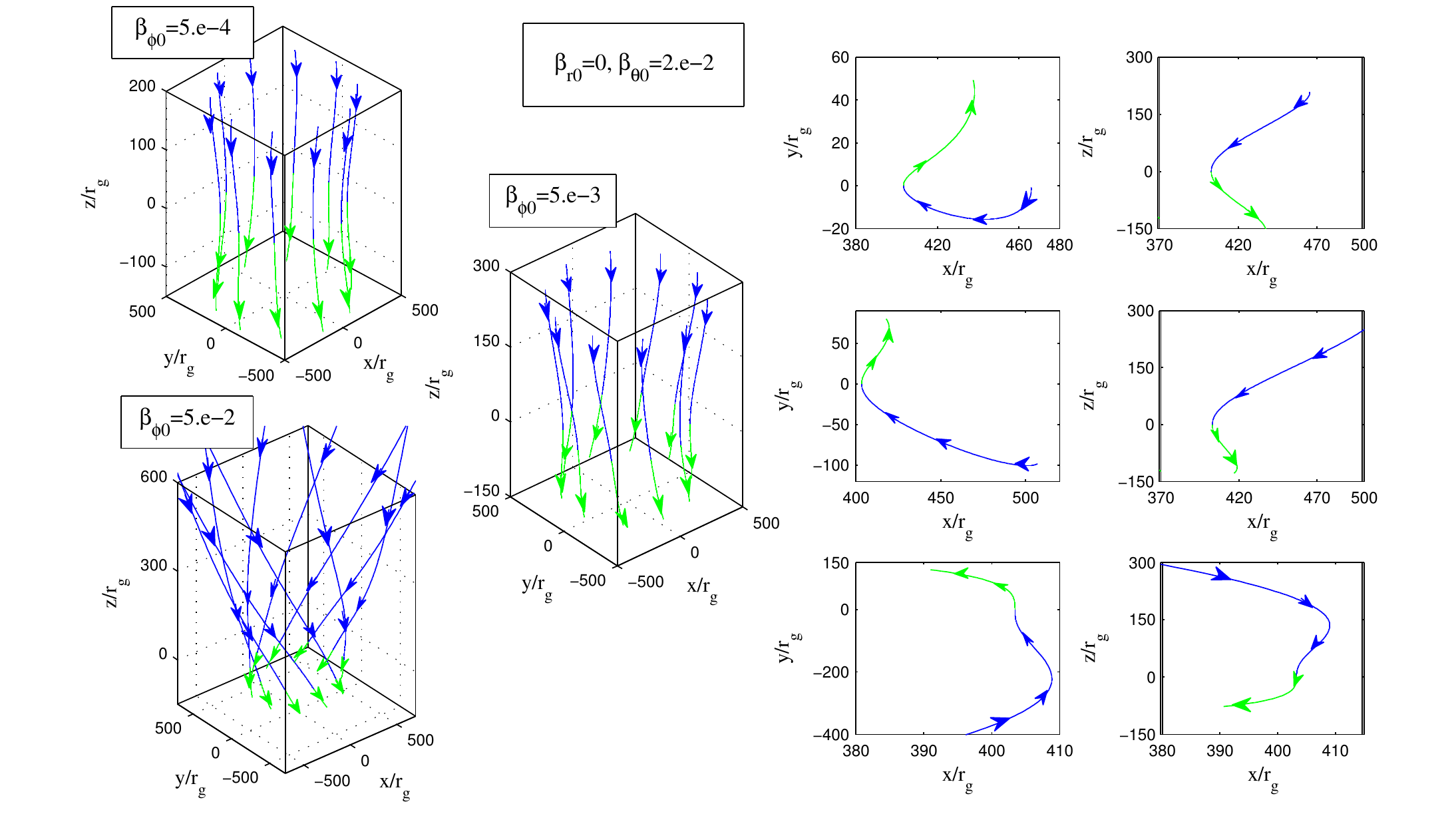}
 \caption{3D presentation of magnetic field lines for 3 different cases which display deviations from even symmetry where $n=0.85, \gamma=\frac{5}{3}, \alpha=\eta_0=0.1$ and $ f=1$. We have fixed $\beta_{r0}=0$ and $\beta_{\theta0}=2\times 10^{-2}$. The 2 columns panels in the right-hand side show the same magnetic lines in $x-y$ plane and in $x-z$ plane. }
\end{figure*}
\begin{figure*}
\centering
\includegraphics[width=170mm]{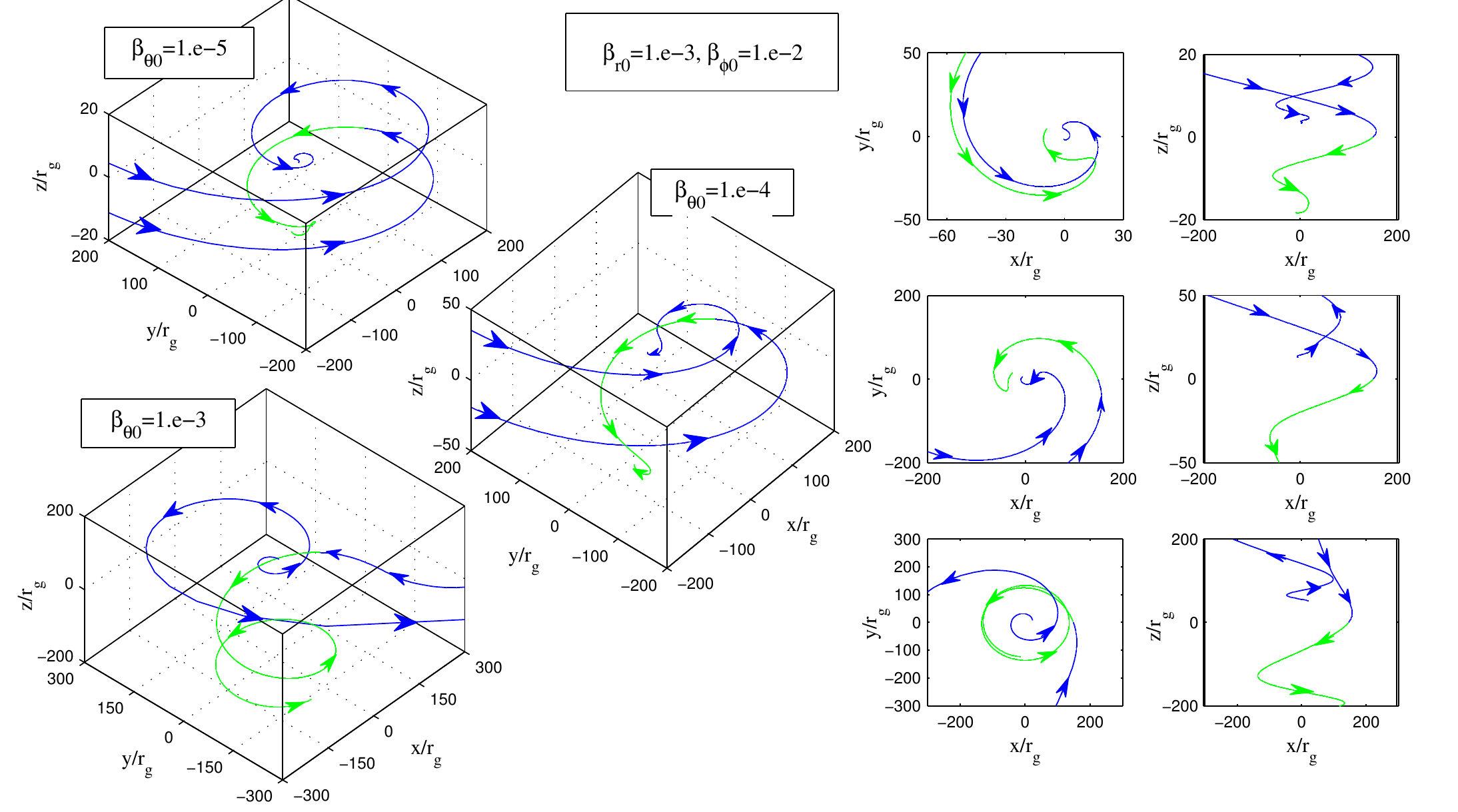}
 \caption{3D presentation of magnetic field lines for 3 different cases which display deviations from odd symmetry where $n=0.85, \gamma=\frac{5}{3}, \alpha=\eta_0=0.1$ and $ f=1$. The 2 columns panels in the right-hand side show the same magnetic lines in $x-y$ plane and in $x-z$ plane.    }
\end{figure*}
\begin{figure*}
\centering
\includegraphics[width=160mm]{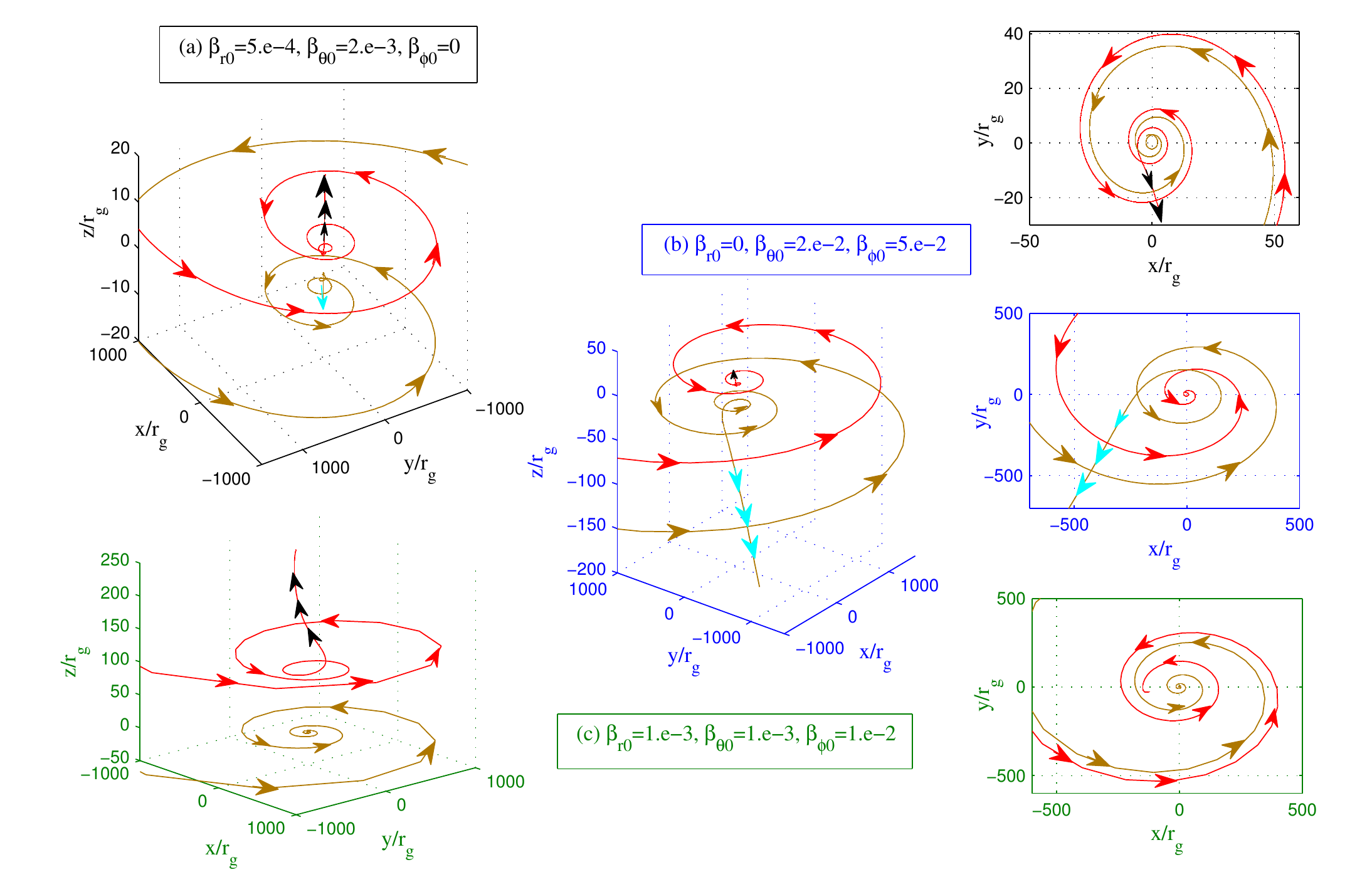}
 \caption{Streamlines for $n=0.85, \gamma=\frac{5}{3}, \alpha=\eta_0=0.1$ and $ f=1$. The red colours show streamlines of northern hemisphere while the brown ones belong to the southern hemisphere of disc. The black arrows displays northern outflows and light blue ones directed outflow in the other side of the equator.  }
\end{figure*}
The last presentation of magnetic field in Fig.12, is devoted to non-symmetric field based on a primary odd one. Here, we can see the same orientation of $B_\phi$ (according to the middle panels). $B_\theta$ becomes zero in a certain angle in the upper side of the disc and it yields a singularity in radius. Moving from the mid-plane along a magnetic line upwards (which is specified with blue line) depicts a counterclockwise path (i.e. $B_\phi>0$) then $r\rightarrow\infty$ (due to $B_\theta\rightarrow 0$), after that singularity the projection continues in the same direction as before but in the opposite vertical direction (because $B_\theta<0$). For all 3 cases of this figure, there is always a "reversal" radius which is the feature of odd symmetrical B-fields (see Fig.4 of Lovelace et al. 1987). In this kind of configuration we can't find 2 different magnetic poles which begin on the north pole and terminate on the south pole (see Fig.1 (b) Wang et al. 1990), it's like unipolar dynamo model of Lovelace (1976) and Blandford (1976) . Every end of a magnetic line terminate to a certain point because in an odd configuration of $\textbf{B}-$ field, $B_\phi$ and $B_r$ don't change their directions in $z=0$ and stay with a constant sign at both sides of the disc.  But in the last row panels of Fig.12, we can specify two magnetic poles whose north is located above the equator. In this case, we had assume $\beta_{\theta0}\sim\beta_{r0}$ and it caused $B_r$ become positive below the midplane which means $B_r$ behaviour is like in the even symmetry although the trend of $B_\phi$ is almost the same as an odd one.
   
\subsection{Asymmetric Stream Lines}
Figure 13 represents the projection of a particle moving parallel to the fluid current. It's seen that the particle begins the path from a very far from the centre (due to $v_\theta\sim 0$ in the mid-plane). Then it moves inwards and along a spiral path which means accretion process is working i.e. $v_r<0$. The less number of cycles in this spiral path means radial velocity is noticeable but not as much as $v_\phi$. In the northern hemisphere, the particle moves downwards at first. Close to the centre, its direction changes to the north and outside the disc (we can call it a reversal angle which is equal opening angle). There is similar spiral path in the southern hemisphere because of mirror symmetry assumption about the mid-plane. For those asymmetrical samples in Fig.11, for case (a) in $x-y$ plane we can see a bit changes in the inflow region, i.e. in every $\theta$ the fraction of $v_\rho/v_\phi$ is almost the same as $\pi-\theta$ (where $v_\rho=v_r\sin\theta+v_\theta\cos\theta$). In case (a), the outflow path is more significant in the upper side (Notice we have shown just a part of outflow region where the total velocity is subsonic). In case (b) and (c), there is just 1 remarkable way out in one side of them. In these 2 cases differences of $v_\rho/v_\phi$ in two sides of disc are more remarkable. In case (c), the path of outflow is almost pure vertically in comparison with (a) and (b) according to the right panels. 

\subsection{Bernoulli Parameter}

\begin{figure*}
\centering
\includegraphics[width=180mm]{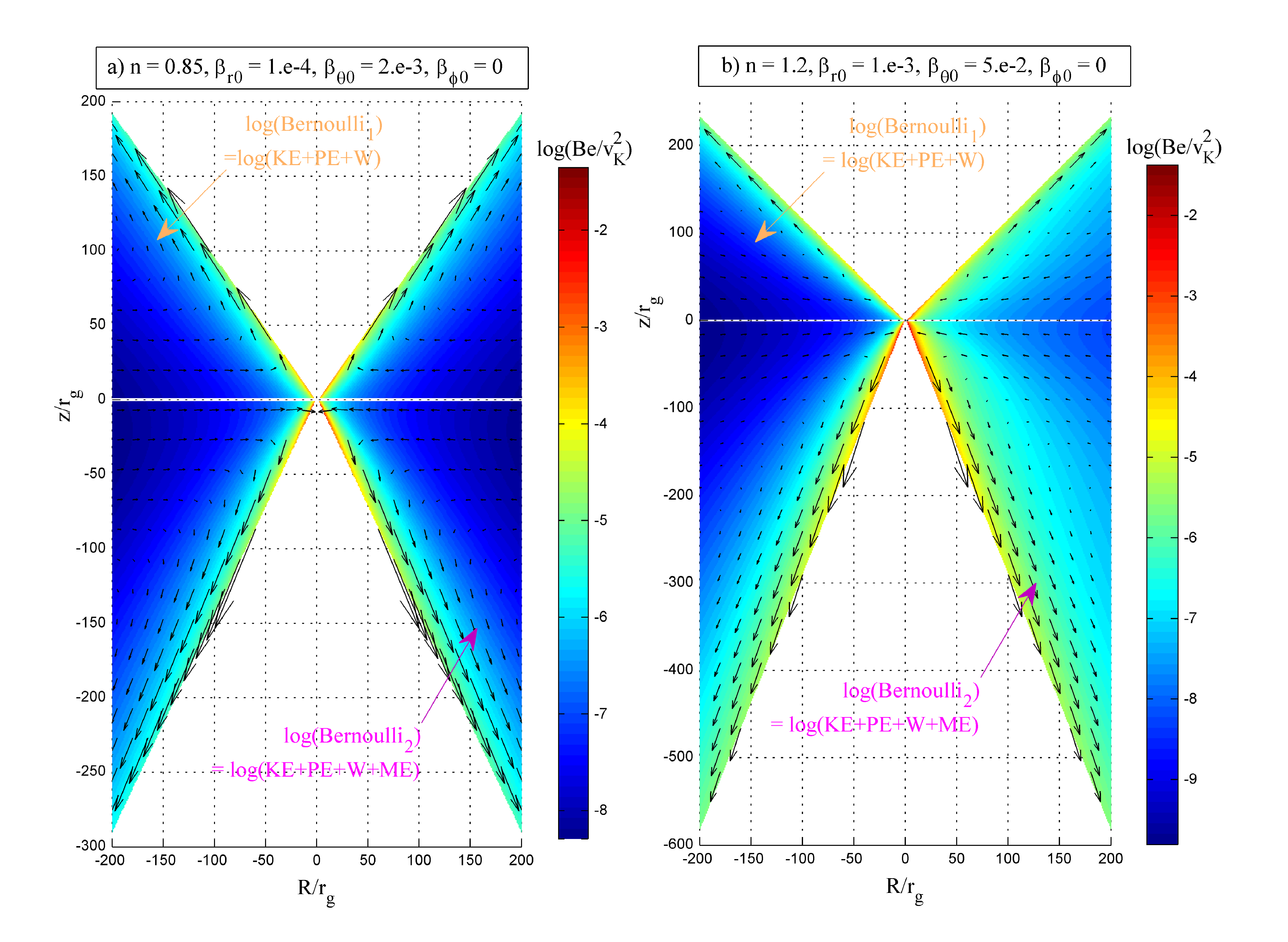}
 \caption{Bernoulli function with two relationships are shown by colours for two different density indexes and magnetic fields. The first formula is summation of kinetic energy ($KE$), potential energy($PE$) and enthalpy ($W$) per unit of mass which have been shown in the left-hand side panels. The right- hand side panels have been plotted with the second formula of Bernoulli function which is similar to the previous one but with an extra term related to magnetic energy ($ME$).  }
\end{figure*}
The Bernoulli function is very important for the study of hydrodynamics outflows. This parameter determines the whole energy per unit mass ,as following
\begin{equation}
Be=\frac{1}{2}v_{tot}^2-\frac{GM_*}{r}+\frac{\gamma}{\gamma-1}\frac{p}{\rho},
\end{equation}
where $v_{tot}$ is the total velocity. When the flow is assumed to be stationary and inviscid without any extra sources or losses, $Be$ is conserved along every streamline. NY95a noticed the positive value of Bernoulli function denoted to the probabilities of outflow presence in their solutions. Moreover, Blandford \& Begelman 1999 had the same opinion and refered it to the origin of the outflow. 
Although a positive value of $Be$ means that the accretion flow can escape to infinity due to its enough energy to overcome
gravitational energy, some researchers pointed out this positive sign of $Be$ is just a result of self-similar solutions (Abramowicz et al. 2000; Yuan 1999).  

Yuan et al. (2012) mentioned the importance of $Be$ and found it positive in the most regions. According to Fukue 1990, for magnetised flow there must be another term in Bernoulli parameter which is related to the magnetic energy. Thus, $Be$ should be written as, 
\begin{equation}
Be_m=\frac{1}{2}v_{tot}^2-\frac{GM_*}{r}+\frac{\gamma}{\gamma-1}\frac{p}{\rho}+\frac{B^2}{4\pi\rho}
\end{equation}
where $B^2$ is the total magnetic field squared. In figure 14, we have shown $Be$ and $Be_m$ by colour. In the left-hand side of this figure, the density index is less than unity, hence the magnetic field is limited to small values. As it's seen the both sides of the left panel of Fig.14 are very similar due to the smallness of magnetic energy. On the other hand, the right-hand side panel shows much more differences between $Be$ and $Be_m$ because of $n>1$ and therefore the larger magnetic field. 
The neglecting of the magnetic term in Bernoulli function might be one reason of finding negative value of $Be$ even in outflow regions in other works like Stone et al. (1999),  Igumenshchev \& Abramowicz (1999) and Yuan \& Bu (2010). Another reason might be the possibility of rejoining outflow to the main accretion disc in further distances (Yuan et al. 2012a).

\subsection{Magnetorotational Instability}
In this section we invistigate the MRI in the asymmetric disc by calculation of the averaged value of magnetic parameters, $<\beta_1>^{-1}=<p/p_m>$ and $<\beta_2>^{-1}=<\rho v^2/p_m>$. Following NY95a, we introduce averaged $\beta_i$ as below,
\begin{equation}
<\beta_i>_{top}=\frac{\int_{\theta_s}^{\pi/2}\beta_i\rho d\theta}{\int_{\theta_s}^{\pi/2}\rho d\theta},	
\end{equation}
\begin{equation}
<\beta_i>_{bottom}=\frac{\int_{\pi/2}^{\pi-\theta_s}\beta_i\rho d\theta}{\int_{\pi/2}^{\pi-\theta_s}\rho d\theta},
\end{equation}
where $\theta_s$ is opening angle. Moreover, the proper condition for MRI is $<\beta_1>^{-1}=<p_g/p_m> \ge 1-100$. As it seen in Tables 1-3, the averaged values of $\beta_1$ are large enough to confirm MRI. In Fig.15 we have presented the inverse magnetic parameter versus polar angle. Local values of $\beta_1$ and $\beta_2$ are much greater than unity except for parts near the disc's surfaces in panels (a) and (b). 
\begin{figure*}
\centering
\includegraphics[width=7in]{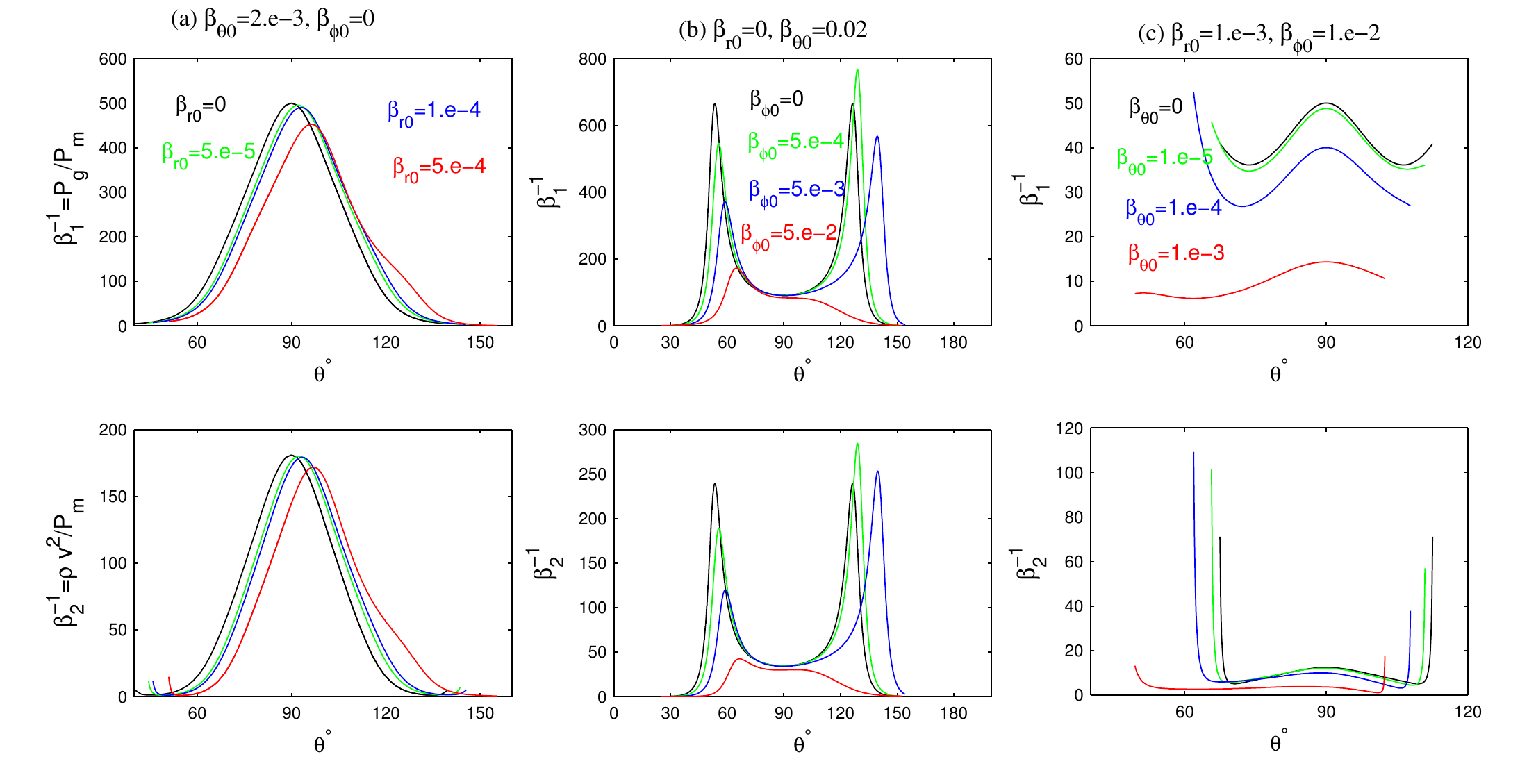}
 \caption{The inversed magnetic parameters versus polar angle }
\end{figure*}
\begin{table}
	\caption{The inverse averaged value of $\beta$'s in each side of the disc
		with $\beta_{\theta0}=2\times 10^{-3}, \beta_{\phi0}=0$ }
	\centering
	\begin{tabular}{l c c rrrr}
		\hline\hline
		$\beta_{r0}$ & Location &$<\beta_1>^{-1}$ & $<\beta_2>^{-1}$ \\[0.5ex] 
		\hline
		&top& 221.96   & 215.00 \\[-1ex]
		\raisebox{1.5ex}{0} \\[1ex]
		& bottom & 221.96   & 215.00\\[-1ex]
		\hline
		&top& 209.14    & 197.46 \\[-0.7ex]
		\raisebox{1.5ex}{5.e-5} \\[1ex]
		& bottom & 234.43  & 230.60  \\[-1ex]
		\hline
		&top& 202.15    & 189.34 \\[-0.7ex]
		\raisebox{1.5ex}{1.e-4}\\[1ex]
		& bottom & 236.69    &235.08 \\[-1ex]
		\hline
		&top& 170.32 & 154.22  \\[-0.7ex]
		\raisebox{1.5ex}{5.e-4} \\[1ex]
		& bottom & 224.77  & 234.53  \\[-1ex]
		\hline
	\end{tabular}
	\label{tab:PPer}
\end{table}
\begin{table}
	\caption{The inverse averaged value of $\beta$'s in each side of the disc
		with $\beta_{r0}=1.\times 10^{-3}, \beta_{\phi0}=1.\times 10^{-2}$ }
	\centering
	\begin{tabular}{l c c rrrr}
		\hline\hline
		$\beta_{\theta0}$ & Location &$<\beta_1>^{-1}$ & $<\beta_2>^{-1}$ \\[0.5ex] 
		\hline
		&top& 179.91   & 182.68 \\[-1ex]
		\raisebox{1.5ex}{0} \\[1ex]
		& bottom & 179.91   & 182.68\\[-1ex]
		\hline
		&top& 163.45    & 163.78\\[-0.7ex]
		\raisebox{1.5ex}{1.e-5} \\[1ex]
		& bottom & 189.97  & 195.62  \\[-1ex]
		\hline
		&top& 135.17    & 131.45 \\[-0.7ex]
		\raisebox{1.5ex}{1.e-4}\\[1ex]
		& bottom & 167.54    &181.00 \\[-1ex]
		\hline
		&top& 84.58 & 71.88  \\[-0.7ex]
		\raisebox{1.5ex}{1.e-3} \\[1ex]
		& bottom & 50.72 & 51.15  \\[-1ex]
		\hline
	\end{tabular}
	\label{tab:PPer}
\end{table}
\begin{table}
	\caption{The inverse averaged value of $\beta$'s in each side of the disc
		with $\beta_{r0}=0, \beta_{\theta0}=2 \times 10^{-2}$ }
	\centering
	\begin{tabular}{l c c rrrr}
		\hline\hline
		$\beta_{\phi0}$ & Location &$<\beta_1>^{-1}$ & $<\beta_2>^{-1}$ \\[0.5ex] 
		\hline
		&top& 41.81  & 31.41 \\[-1ex]
		\raisebox{1.5ex}{0} \\[1ex]
		& bottom & 41.81  & 31.41\\[-1ex]
		\hline
		&top& 40.63  & 31.13\\[-0.7ex]
		\raisebox{1.5ex}{5.e-4} \\[1ex]
		& bottom & 41.01  & 30.30  \\[-1ex]
		\hline
		&top& 32.98    & 26.69 \\[-0.7ex]
		\raisebox{1.5ex}{5.e-3}\\[1ex]
		& bottom & 34.36 &24.71 \\[-1ex]
		\hline
		&top& 9.26 & 9.10  \\[-0.7ex]
		\raisebox{1.5ex}{5.e-2} \\[1ex]
		& bottom & 13.09 & 9.01  \\[-1ex]
		\hline
	\end{tabular}
	\label{tab:PPer}
\end{table}

\section{Conclusion}
We studied an asymmetric configuration of a magnetic field in self-similar approach. We assumed a small deviation from an odd or even symmetrical field which are introduced by Lovelace et al. (1986). According to the even configuration, both radial and azimuthal components of \textbf{B} are zero in the equator ($\theta=90^\circ$) while $B_\theta|_{90}\neq 0$. We examined once $B_{r0}\neq 0$ and another time $B_{\phi0}\neq0$ (the two other components were the same as the previous even case). We have presented the results of self-similar solutions in the radial direction and found a set of ODE's with respect to polar angle. By taking into account the reflection symmetry about the equatorial plane we could solve numerically the differential equations. Our results showed the half-thickness of upper side isn't equal the other side one. We saw the upper side's half-thickness is greater when $\beta_{\phi0}=B_{\phi0}^2/(8\pi p_0)$ is dominant.  On the other hand, when $\beta_{r0}\neq 0$ (we assumed $B_{r0}<0$), the disc shrinks above the mid-plane and expands below. 
We found out in the case of density parameter less than unity, the magnetic forces can not have a positive effect directly on driving outflows. On the other hand, the straightforward positive radial acceleration is mainly due to the gradient of gas pressure which gives rise to outflows. The outflow from one side or even two sides of the disc take noticeable fraction of kinetic energy and mass outwards. 

Breaking reflection symmetry about equatorial plane from disc' outflows is not easy to see like asymmetrical shape in jets and counter jets from some AGNs. Jets are directly observable with shining ionised gas propagating at nearly the speed of light but outflows with smaller velocities ($v\leq c/3$)  are known from the spectra of AGNs. In this work, we have clearly pointed out that the complex behaviour of the accretion flow depends on the input parameters. Consideration of the space outside the disc is the most important improvement of this study and it provides a better interpretation for spectrum of AGNs with intrinsically one-sided jets.

\section*{Acknowledgment}
SA acknowledges support from department of astronomy, Cornell university for their hospitality during a visit which a part of this work has been done there. We also appreciate the referee for his/her thoughtful and constructive comments on an early version of the paper.

\appendix
\section{}
 In the spherical coordinates, by the assumption of axisymmetric and steady state, $\partial/\partial\phi=0$, $\partial/\partial t=0$, the continuity and three components of momentum equation can be respectively written as:
\begin{equation}
\frac{1}{r^2}\frac{\partial}{\partial r}(r^2 \rho v_r)+\frac{1}{r \sin\theta}\frac{\partial}{\partial \theta}(\sin\theta \rho v_\theta)=0,
\end{equation}
\begin{displaymath}
    v_r\frac{\partial v_r}{\partial r}+\frac{v_\theta}{r}\frac{\partial v_r}{\partial\theta}-\frac{1}{r}(v_\theta^2+ v_\phi^2)=-\frac{GM}{r^2}-\frac{1}{\rho}\frac{\partial p}{\partial r}
  \end{displaymath}
   \begin{equation}
  \hspace{1cm}-\frac{1}{4\pi\rho r}\{B_\phi\frac{\partial}{\partial r}(rB_\phi)+B_\theta[\frac{\partial}{\partial r}(rB_\theta)-\frac{\partial B_r}{\partial \theta}]\},
\end{equation}
\begin{displaymath}
  v_r\frac{\partial v_\theta}{\partial r}+\frac{v_\theta}{r}\frac{\partial v_\theta}{\partial\theta}+\frac{1}{r}(v_r v_\theta-v_\phi^2\cot\theta)=-\frac{1}{\rho r }\frac{\partial p}{\partial \theta}
    \end{displaymath}
  \begin{equation}
\hspace{1cm}+\frac{1}{4\pi\rho r}\{B_r[\frac{\partial}{\partial r}(rB_\theta)-\frac{\partial B_r}{\partial\theta}]-\frac{B_\phi}{\sin\theta}\frac{\partial}{\partial\theta}(B_\phi \sin\theta)\},
\end{equation}
\begin{displaymath}
v_r\frac{\partial v_\phi}{\partial r}+\frac{v_\theta}{r}\frac{\partial v_\phi}{\partial\theta}+\frac{v_\phi}{r}(v_r+v_\theta \cot\theta)
=\frac{1}{\rho r^3}\frac{\partial}{\partial r}(r^3 T_{r\phi})
\end{displaymath}
 \begin{equation}
\hspace{1cm}+\frac{1}{4\pi\rho r}[B_r\frac{\partial}{\partial r}(rB_\phi)+\frac{B_\theta}{\sin\theta}\frac{\partial}{\partial\theta}(B_\phi \sin\theta)],
\end{equation}
The energy equation becomes:
\begin{equation}
\rho(v_r\frac{\partial e}{\partial r}+\frac{v_\theta}{r}\frac{\partial e}{\partial\theta})
-\frac{p}{\rho}(v_r\frac{\partial\rho}{\partial r}+\frac{v_\theta}{r}\frac{\partial\rho}{\partial\theta})=f [T_{r\phi} r\frac{\partial}{\partial r}(\frac{v_\phi}{r})]
\end{equation}
  Since we assume the steady flow then the left hand side of the induction equation becomes zero, $\partial \textbf{B}/\partial t=0,$  so we have
\begin{equation}
 \frac{1}{r^2\sin\theta}\frac{\partial}{\partial\theta}[\sin\theta\{r(v_rB_\theta-v_\theta B_r)-\eta (\frac{\partial(rB_\theta)}{\partial r}-\frac{\partial B_r}{\partial\theta})\}]=0,
\end{equation}
\begin{equation}
 -\frac{1}{r\sin\theta}\frac{\partial}{\partial r}[\sin\theta\{r(v_rB_\theta-v_\theta B_r)-\eta (\frac{\partial(rB_\theta)}{\partial r}-\frac{\partial B_r}{\partial\theta})\}]=0,
\end{equation}
\begin{displaymath}
\frac{1}{r}\{\frac{\partial}{\partial r}[r(v_\phi B_r- v_r B_\phi)+\eta\frac{\partial(r B_\phi)}{\partial r}]
+\frac{\partial}{\partial\theta}[(v_\phi B_\theta-v_\theta B_\phi)
\end{displaymath}
\begin{equation}
\hspace{1cm}+\frac{\eta}{r\sin\theta}\frac{\partial}{\partial\theta}(\sin\theta B_\phi)]\}=0.
\end{equation}
Comparing (A6) and (A7) leads us to:
\begin{equation}
\frac{\partial(rB_\theta)}{\partial r}-\frac{\partial B_r}{\partial\theta}=\frac{r}{\eta}(v_rB_\theta-v_\theta B_r),
\end{equation}

\begin{equation}
\nu=\frac{\alpha r}{\rho v_K}(p+\frac{B^2}{8\pi}),
\end{equation}

\begin{equation}
\eta=\frac{\eta_0 r}{\rho v_K}(p+\frac{B^2}{8\pi}),
\end{equation}
where $\eta_0$ is a constant value.

\end{document}